\documentclass[twocolumn,nofootinbib]{revtex4-1}
\usepackage{verbatim}
\usepackage{graphicx}
\usepackage{float}
\usepackage{amssymb}
\usepackage[utf8]{inputenc}
\usepackage[english]{babel}
\usepackage{amsmath}
\usepackage{mathtools}
\usepackage{amsfonts}
\usepackage{stackrel}
\usepackage{cases}
\usepackage{bm}%
\usepackage[export]{adjustbox}
\usepackage{hyperref}

\newcommand{\be}{\begin{align}}
\newcommand{\ee}{\end{align}}
\newcommand{\dif}{\mathrm{d}}

 \newcommand{\abs}[1]{\left\vert#1\right\vert}
 \newcommand{\defeq}{\stackrel{\text{def}}=}
  \usepackage{color}

\definecolor{blue}{rgb}{0,0,1}
\definecolor{green}{rgb}{0,1,0}
\definecolor{red}{rgb}{1,0,0}
\definecolor{vio}{rgb}{1,0,1}
\definecolor{uv}{rgb}{0.5,0,0.5}
\definecolor{ama}{rgb}{0.3,0.3,0.3}

\begin{document}
 \title{Genuine localisation transition in a long-range hopping model}
 \author{Xiangyu Cao}\author{Alberto Rosso}\affiliation{LPTMS, CNRS, Univ. Paris-Sud, Université Paris-Saclay, 91405 Orsay, France}\author{Jean-Philippe Bouchaud}\affiliation{Capital Fund Management, 23 rue de l'Universit\'e, 75 007 Paris, France}\author{Pierre Le Doussal}\affiliation{CNRS-Laboratoire de Physique Théorique de l'\'Ecole Normale Supérieure, 24 rue Lhomond, 75231 Paris Cedex, France}
 \begin{abstract}
 	We introduce and study a new class of Banded Random Matrix model describing sparse, long range quantum hopping in one dimension. Using a series of analytic arguments, numerical simulations, and mappings to statistical physics models, we establish the phase diagram of the model. A genuine localisation transition, with well defined mobility edges, appears as the hopping rate decreases slower than $\ell^{-2}$, where $\ell$ is the distance. Correspondingly, the decay of the localised states evolves from a standard exponential shape to a stretched exponential and finally to a novel $\exp(-C\ln^\kappa \ell)$ behaviour, with $\kappa > 1$.
 \end{abstract}
 \maketitle 
 
 \section{Introduction}\label{sec:intro}
 Since the premonitory work of Anderson \cite{anderson1958absence}, the existence of localisation transitions of eigenstates of random Hamiltonians is a fundamental and non-trivial problem. It was settled for the single-particle Anderson model in $d$-dimension only much later \cite{abraham79scaling}. When $d > 2$ the localisation transition exists and manifests itself as \textit{mobility edges} in the eigenvalue spectrum, that separate extended and localised eigenstates in the spectrum. At the vicinity of the mobility edge, the eigenstates have multi-fractal statistics, described by a localisation transition critical point. When $d \leq 2$, on the other hand, all eigenstates are localised (for the standard Anderson model) and there is no mobility edge. There has been a renewed surge of interest on these issues in the context of many-body localisation (see \textit{e.g.} 
 \cite{Basko2006mbl,huse15mblreview,altman15mbl}), with intriguing suggestions of the possibility of extended, but non ergodic quantum states.

 In this respect, $1$-d random Hamiltonians with long-range hopping are important laboratories to understand localisation transitions, as they can be seen as proxies for higher 
 dimension models. The best-known such model is the Power-law Random Banded Matrix (PBRM) ensemble \cite{mirlin96pbrm}. Its elements $H_{nm}$ are independent centered Gaussian random variables such that \footnote{Our notation is related to that in the literature \cite{mirlin96pbrm} by $\mu = \sigma = 2 \alpha - 1 > 0$ and $\beta = b^{-\mu-1} > 0$.} 
 \begin{eqnarray}
 \overline{H_{nm}^2}^c = \frac{1}{1 + g_{nm}}\, ;\quad g_{nm} := \beta \abs{n-m}^{1+\mu}, \label{eq:pbrmasymp}  
 \end{eqnarray}
 where $\beta, \mu > 0$ are some coefficients. The PBRM phase diagram can be qualitatively understood by comparing the typical nearest level spacing $\sim 1/N$ (where $N$ is the size of the matrix) with the \textit{direct} hopping element $\gtrsim 1/N^{\frac{1 + \mu}{2}}$. When $\mu < 1$,  all the eigenstates are extended; when $\mu > 1$, they are all localised but with power-law decay \cite{mirlin96pbrm}.  Remarkably, at $\mu = 1$, there is a line of critical models (parametrised by $\beta$) with multi-fractal eigenstates, which were studied numerically \cite{cuevas2001anomalously} and analytically \cite{levitov99pbrm,mirlin00pbrm} (see \cite{evers2008anderson} for a review, and \cite{quito2016anderson,amini17spread} for recent work). Yet, there is no mobility edge in this case, making this transition qualitatively different from the conventional localisation transition. The phase diagram is oversimplified because \textit{direct} tunneling dominates transport in the quantum \textit{small world} that PBRM describes. In fact, the above picture applies whenever the entry distributions are of the form $P(H) = \sqrt{g} P_0(\sqrt{g} H)$ for $g \to \infty$, where  $P_0$ is ``narrow'' such that the moments obey $\overline{H^p} \propto g^{-p/2}$ in the large $g$ limit. This means that all matrix elements corresponding to a certain (large) distance $|n-m|$ have the same order of magnitude.   
 
 In this work, we define a new class of Banded Random Matrices, which we call \textit{broadly distributed}, and composed by \textit{quasi--sparse} matrices. Its matrix elements $Q_{mn}$ have the following probability distribution: 
  \begin{equation} 
  \abs{Q_{mn}} \sim \begin{dcases}
  O(1) & \text{with prob. $1/g_{mn}$} \\
  0 & \text{otherwise.} 
  \end{dcases} \label{eq:Qmndef0}
  \end{equation}
  The typical elements are very close to zero while few ``black swans'' are\textit{ random} numbers of order $1$. 
  Two representative models will be studied with care and defined in section \ref{sec:definition}: (i) the randomised sparse matrix (RSM) model, and (ii) the Beta banded random matrix (BBRM),  which enjoys an exact mapping with a long--range epidemic model studied recently~\cite{hallatschek2014acceleration,chatterjee2016multiple}.
  
\begin{figure}
 \center
 \includegraphics[width=.9\columnwidth]{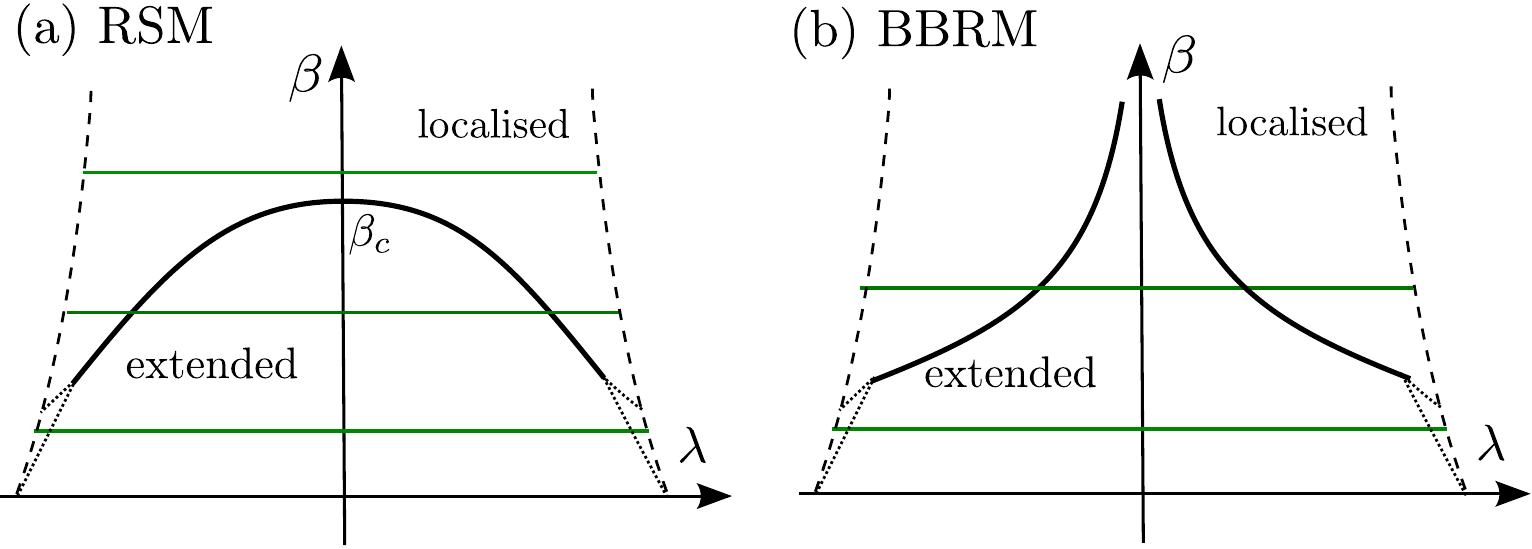}
 \caption{(a) A qualitative sketch of the phase diagram of the RSM in the regime $\mu \in (0, 1)$, in the parameter plane of the disorder strength $\beta$ and the energy $\lambda$. (b) The analogous phase diagram for the BBRM ensemble. It differs from (a) in that $\beta_c = \infty$: mobility edges always for any large $\beta$. }\label{fig:sparsephase} 
 \end{figure}

 We shall argue, analytically and numerically, that for any random banded matrix model in the broadly distributed class, and for any \textit{fixed} $\mu \in (0, 1)$, there exists a localisation transition with generically mobility edges. The phase diagram of the broadly distributed class in the $\mu \in (0,1)$ regime is thus different from PBRM, and is illustrated in Figure \ref{fig:sparsephase}. 
 
 The rest of the paper is organised as follows. Section \ref{sec:definition} defines the general broadly distributed class, and introduces its two representatives, the BBRM and the RSM. Section \ref{sec:dos} focuses on the density of states of these models, as a preparation of the localisation properties. Section \ref{sec:LT} presents the analytical argument (section \ref{sec:block}) and numerical evidence (section \ref{sec:num}) supporting the localisation transition in the $\mu \in (0, 1)$ regime.  In section \ref{sec:decay}, we describe the spatial decay of the localised states for $\mu > 0$, and discuss the effective dimension of the model as a function of $\mu$. Our predictions are based on the mapping with two statistical physics models: the epidemics model (section \ref{sec:fisher}) and the long--range percolation model (section \ref{sec:LRP}). 
 
 \section{Broadly distributed random banded matrices}\label{sec:definition}
 \subsection{General definition}
 We define now the broadly distributed class. Throughout, we still consider real symmetric random matrices $Q_{mn} = Q_{nm}$ ($m,n=1, \dots, N$), whose matrix elements are otherwise uncorrelated. Their probability distribution depends only on the distance \footnote{In numerical simulations, periodic boundary conditions are always used. The distance between sites will be modified as $\abs{m-n}\leadsto N - \abs{m-n}$ for $\abs{m-n}>N/2$.} from diagonal through the variable 
\begin{equation} g = \beta \abs{m-n}^{\mu + 1} \,;  \label{eq:g}\end{equation}
  Note that for simplicity we shall drop the subscript: $g := g_{mn}$ with respect to Eq. \eqref{eq:pbrmasymp}. As mentioned in section \ref{sec:intro}, the matrix $Q_{mn}$ is quasi--sparse, with   typical matrix elements exponentially small:
 \begin{equation}
 \abs{Q_{mn}}^{\text{typ.}} \leq e^{-cg} \,. \label{eq:Qtyp}
 \end{equation}
 More precisely, we require that the cumulant generating function of $\abs{Q_{mn}}$ satisfies the following:
   \begin{subequations}\label{eq:broaddefall}
   \begin{equation}
   \ln \overline{e^{-t \abs{Q_{mn}}}} \stackrel{g\gg1}=  - \frac1g f(t) + O(g^{-2})  \,,\label{eq:broaddef}
   \end{equation}
  where $f(t)$ is some function \textit{independent} of $g$ and \textit{slowly varying}: $\lim_{t\to +\infty} f(at) / f(t) = 1$ for all $a > 0$. For convenience, we shall work with the assumption that $f$ grows at most logarithmically:
   \begin{equation} f(t \rightarrow  +\infty) = O(\ln t) \,. \label{eq:ftlog}\end{equation}
     \end{subequations}
   As we show in more detail in Appendix \ref{sec:math1}, equations Eq. \eqref{eq:broaddefall} imply the quasi--sparseness properties Eq. \eqref{eq:Qmndef0}. More precisely Eq. \eqref{eq:ftlog} guarantees that the $\abs{Q_{mn}}^{\text{typ.}} \lesssim e^{-cg}$, while Eq. \eqref{eq:broaddef} corresponds to the black swans, and implies that the moments of the matrix elements satisfy
   \begin{equation}
    \overline{\abs{Q_{mn}}^k} \stackrel{g\gg1}\sim  \frac{c(k)}{g} \label{eq:moments}
   \end{equation} 
   with $c_k$ given by the series expansion $ f(t) = \sum_{k>0} (-1)^{k-1} c(k) t^k / k!$.   All the moments are all dominated by the black swans. In particular, the mean of $\abs{Q_{mn}}$ much larger than its typical value $\abs{Q_{mn}}^{\text{typ.}} \lesssim e^{-g}$.    
   The second moment $\overline{Q_{mn}^2}$ has the same asymptotics as $\overline{H_{mn}^2}$ of PBRM, see Eq. \eqref{eq:pbrmasymp}. However, for $k > 2$, $\overline{\abs{H_{mn}}^k} \sim g^{-k/2}$ is qualitatively different from Eq. \eqref{eq:moments}. 
      
\begin{figure}
    (a)\includegraphics[width=.9\columnwidth,valign=t]{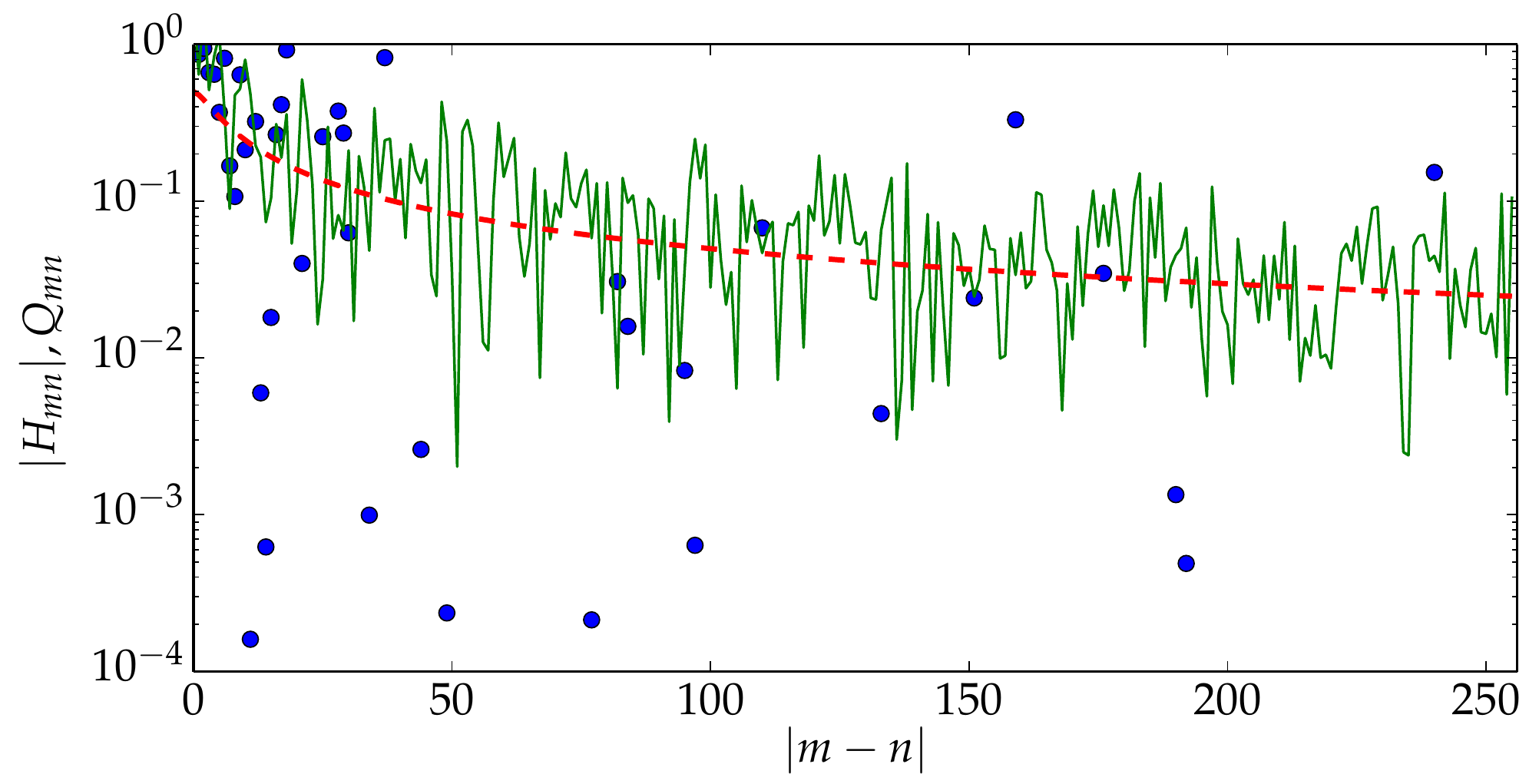} \\
    (b)\includegraphics[width=.39\columnwidth,valign=t]{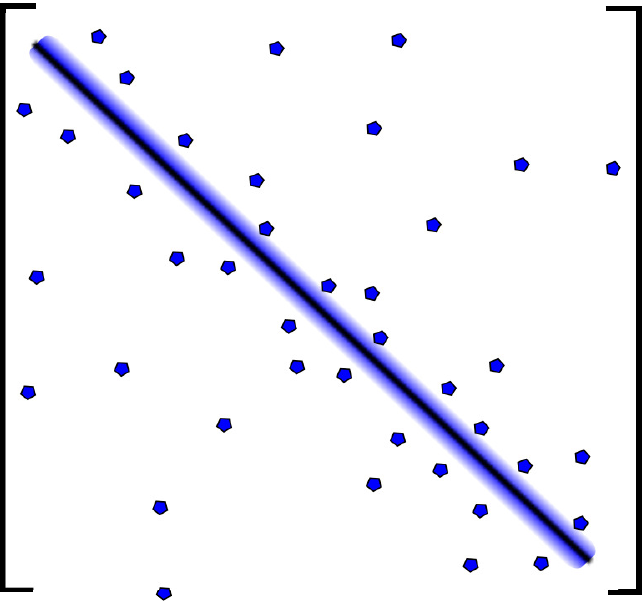} 
    (c)\includegraphics[width=.39\columnwidth,valign=t]{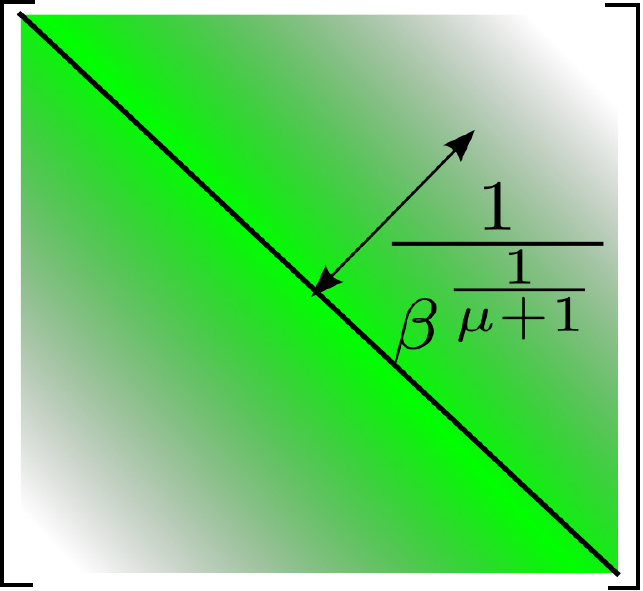} 
    \caption{(a) The absolute value of matrix elements of  PBRM (line) and  BBRM (dots) on one line of the matrix,as a function of the distance from the diagonal. For both matrices, we used the parameters $\mu = .5, \beta = 0.1$. The red dashed line depicts the standard deviation common to both. Many elements of the BBRM are too small to be drawn on the plot. (b) A schematic colour plot of the matrix elements' amplitudes of a broadly distributed banded random matrix. Off--diagonal, the matrix is quasi--sparse. (c) The same plot for the PBRM. The hopping elements have narrowly distributed amplitudes.}  \label{fig:matrixele}
    \end{figure}

   A generic feature of broadly distributed matrices is that the log of $\abs{Q_{nm}}$ has much larger fluctuations than those of $\abs{H_{nm}}$. An illustration is given in Figure \ref{fig:matrixele} (a) using the BBRM ensemble defined below. Such a broad distribution of matrix elements makes this class akin to L\'evy matrices \cite{cizeau1994levy,tarquini16levy}. The latter model is an fully--connected mean field model, and can be studied by cavity methods, as the Bethe lattice Anderson model. This cannot be said for the broadly distributed models, which all retain a non--trivial spatial structure, as we will explore below.
   
   Note that the definition Eq. \eqref{eq:broaddefall} does not specify the precise form of $P(Q_{mn})$; neither does it concern the elements near the diagonal. This leads to a considerable diversity of the broadly distributed class. To illustrate that, two examples will be introduced in the following sections. 
   
   \subsection{Randomised Sparse Matrices}\label{sec:SPdef}
    In the RSM ensemble, the distribution of the off--diagonal elements $Q_{mn}$ is of the form:
    \begin{equation}\label{eq:sparse}
    P(Q_{mn}) = g^{-1} P_0(Q_{mn}) + \left(1 - g^{-1}\right) \delta(Q_{mn}) \,,\,
    \end{equation}
   where $P_0(x)$ is the pdf of a fixed, narrow distribution (\textit{e.g.}, Gaussian or uniform). That is, $Q_{mn} = 0$ with probability $1-1/g$, and is an $O(1)$ random variable with fixed distribution otherwise. 
   
  The randomised sparse matrices are clearly in the broadly distributed class. Indeed, one can show that Eq. \eqref{eq:sparse} satisfied Eq. \eqref{eq:broaddef} with
  $f(t) =  \int \dif v (1-e^{-t \abs{v}}) P_0(v)  \leq 1$, which fulfils Eq. \eqref{eq:ftlog}. 
  
  In the numerical study (section \ref{sec:num}), the distribution $P_0$ is uniform in $[-\frac12, \frac12]$, and the diagonal and nearest neighbour hopping elements ($Q_{mn}$ for $\abs{m-n} \leq 1$) have the uniform distribution $P_0$.

   \subsection{Beta Banded Random Matrices}\label{sec:bbrmdef}
    This BBRM ensemble has vanishing diagonal elements: $Q_{nn} = 0$, while the the off--diagonal elements $Q_{mn}$ are random variable between $0$ and $1$. They obey a special case of the Beta distribution:
    \begin{equation}
      \mathbb{P}(Q_{mn} < q) = q^{1/g} \,,\,  0\leq q  \leq 1 \,, \label{eq:Qdefsupp1}
    \end{equation}
     Equivalently, one may define $Q_{mn}$ as  a ``Boltzmann weight'' (with inverse temperature $\beta$)
     \begin{equation}
     Q_{mn} = e^{-\beta \tau_{mn}} \,,\, P(\tau_{mn}) = \exp\left(-\frac{\tau_{mn}}{\abs{m-n}^{\mu+1}} \right) \,,\, \label{eq:Qdef2}
     \end{equation}
     \textit{i.e.}, $\tau_{mn}$ is an exponential random variables with average  $\abs{m-n}^{\mu+1}$. As we shall see in section \ref{sec:fisher}, the form Eq. \eqref{eq:Qdef2} is directly motivated by the mapping to statistical models. 
     
     BBRM belongs to the broadly distributed class since the definition Eq. \eqref{eq:Qdefsupp1} satisfies the general definition Eq. \eqref{eq:broaddefall}, with
 $ f (t)= \log (t)+\Gamma (0,t)+\gamma_E$,
    which satisfies the bound Eq. \eqref{eq:ftlog} ($\Gamma(x,t)$ is the incomplete Gamma function and $\gamma_E$ is the Euler constant). Note that the BBRM matrix is quasi--sparse but not sparse.

    
 \section{Density of states}\label{sec:dos}
 
 The general definition of the DoS for any random matrix is:
  \begin{equation}
  \rho(\lambda) = \frac1N \sum_{i=1}^N \overline{\delta(\lambda - \lambda_i)} \,,
  \end{equation}
 where $\lambda_1, \dots, \lambda_N$ are the eigenvalues of the matrix. When averaged over disorder, $\rho(\lambda)$ is usually a continuous function that vanishes outside some finite interval $[\lambda_{-}, \lambda_{+}]$ (a notable exception is the Lévy matrix \cite{cizeau1994levy,tarquini16levy,monthus16levy}, for which $\lambda_{+} = + \infty$). 
 As an example, we recall  that the DoS of the PBRM is always given by Wigner's semi--circle law \cite{mirlin96pbrm} (upon a rescaling).
 In general, the width of the support of the DoS can be estimated by 
 $$ \overline{\lambda^2} \defeq \int \rho(\lambda) \lambda^2 \dif \lambda = \frac1N \sum_{i} \overline{\lambda_i^2} =  \frac1N\overline{\mathrm{Tr}\hat{H}^2} =  \frac1N \sum_{nm} \overline{H_{mn}^2}$$
 For both PBRM and the broadly distributed class, the last quantity $\sim \sum_{n=1}^N \beta^{-1} \abs{n}^{-\mu - 1}$ (by Eq. \eqref{eq:moments} and Eq. \eqref{eq:pbrmasymp}). So, when $\mu < 0$, $\overline{\lambda^2} \sim N^{-\mu}$ diverges as $N \to \infty$. In such cases, according to a theorem of \cite{kus1991density}, the DoS must also be a semicircle law for the whole broadly distributed class. 
 \begin{figure}
  \center \includegraphics[width=.49\columnwidth,valign=t]{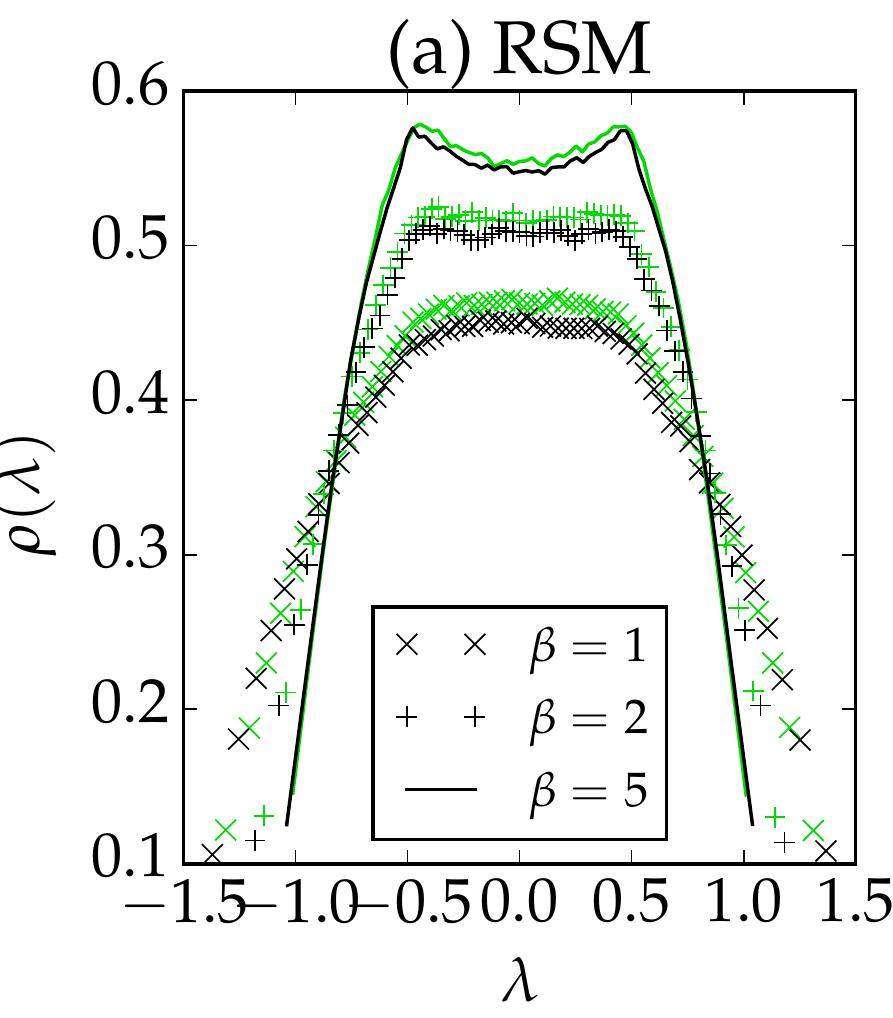} \includegraphics[width=.49\columnwidth,valign=t]{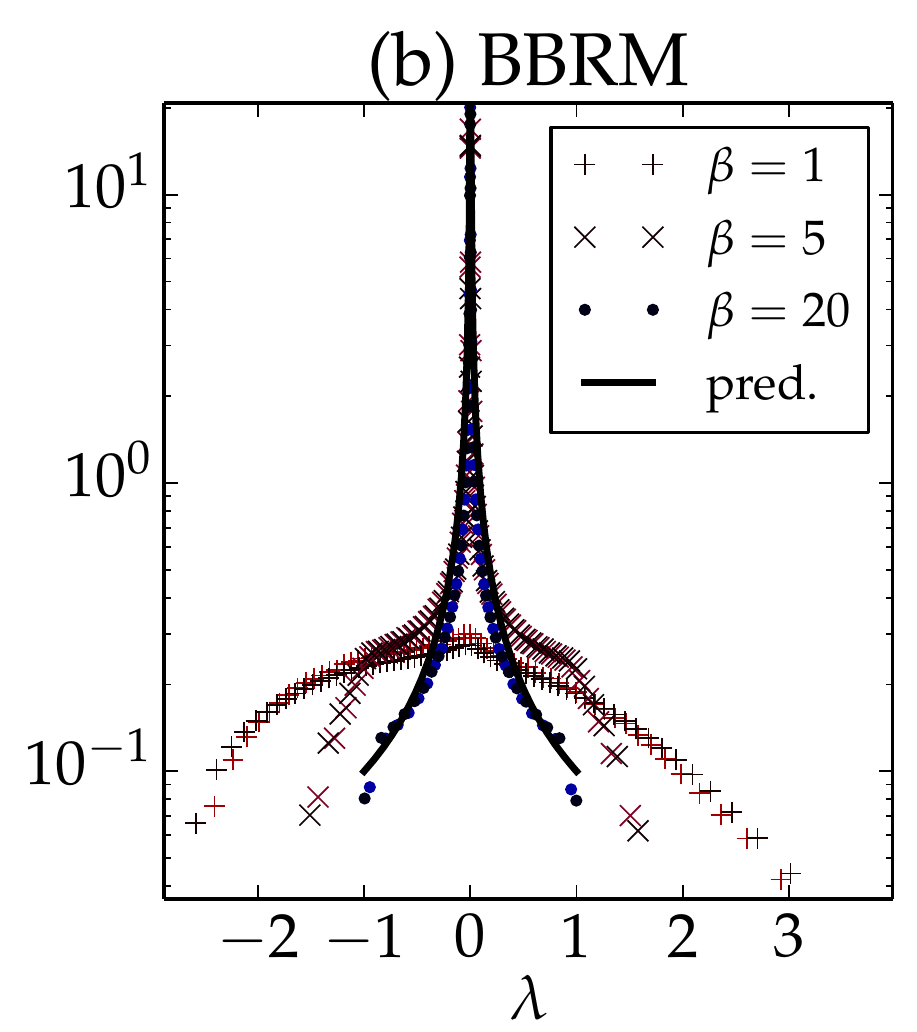}
 \caption{(a) The DoS of the randomised sparse matrix with parameters $\mu = 0.5$, $\beta = 1, 2$ and $5$, and matrix sizes $N = 2^7$ and $2^{13}$. (b) DoS of the BBRM for $\mu = .5$ and various $\beta$. The curve $\beta = 20$ is compared to the prediction Eq. \eqref{eq:rhobigb}, plotted in thick curve.  In all numerical data, lighter colours represent a small system size $N = 2^7$ and the colour black represents a large system size $N = 2^{13}$.}\label{fig:DoS}
 \end{figure}
  
 When $\mu > 0$, the DoS of the broadly distributed class is non--semicircular. Moreover, the DoS is sensible to the matrix elements near the diagonal, so can change qualitatively from one model to another. To illustrate this, we consider numerically the two examples introduced in section \ref{sec:definition}. 
  
 For the randomised sparse matrix, see Figure \ref{fig:DoS} (a), $\rho(\lambda)$ is symmetric  with respect to $\lambda=0$. It is roughly constant in the interval $[-\frac12, \frac12]$ for the range of $\beta \in [1, 5]$ plotted. As $\beta$ increases further, two maxima are displayed at $\lambda = \pm \frac12$, similarly to the DoS of 1D Anderson model. 
 
 The DoS is more singular for the BBRM model. Indeed, see Figure \ref{fig:DoS} (b), $\rho(\lambda)$ develops a divergence at $\lambda = 0$, and a non-analyticity at $\lambda = \pm 1$ as $\beta$ increases. For $\beta \gg 1$, most elements are vanishing, and a crude estimate of DoS is given by the distribution of the eigenvalues $\pm Q_{01} $ of the $2\times2$ sub-matrix $\begin{pmatrix} 0 & Q_{01}  \\ Q_{01} & 0 \end{pmatrix}$. This 
  leads to:
  \begin{equation} 
  \rho(\lambda) \approx \abs{\lambda}^{-1 + 1/\beta} / (2\beta) \,,\,\text{ if } \abs{\lambda} < 1 \label{eq:rhobigb}
  \end{equation} 
  and $\rho(\lambda) \approx 0$ for $\abs{\lambda}> 1$.
  
 By considering other variants of BBRM and the randomised sparse matrix with different definition near the diagonal, we observe the following general pattern: the DoS is singular at $\lambda = 0$ if and only if the near diagonal elements are also quasi--sparse, as is the case for BBRM at large $\beta$. As we shall see below, such a divergence affects the localisation property of states near $\lambda = 0$ and entails a modification of the phase diagram, as shown in Figure \ref{fig:sparsephase}.

 \section{Localisation transition}\label{sec:LT}
 Let us consider a general banded matrix $Q_{mn}$ with broadly distributed elements in the sense of eqs. \eqref{eq:broaddefall}.  Since the typical matrix elements are smaller compared to PBRM, it is natural to expect that the broadly distributed matrices are more localised than PRBM. Indeed, when $\mu > 1$, all the states are localised, as for the PBRM model. However, the decay of the localised states are qualitatively different from PBRM. On the other hand, when $\mu < 0$, all the states are extended. We shall come back to these claims in section \ref{sec:decay}. 
 
 In the rest of section, we concentrate on the regime $\mu \in (0,1)$. In section \ref{sec:block}, we use a block diagonalisation argument to show that there is a localisation transition by fixing the eigenvalue $\lambda$, and tuning the disorder strength $\beta$, \textit{i.e.}, moving vertically in the parameter plane of Fig. \ref{fig:sparsephase}. Then we will argue for the generic existence of mobility edges, corresponding to moving horizontally in  Fig. \ref{fig:sparsephase}. We further confirm this claim by the numerical study of the representative models in section \ref{sec:num}. 
  
 \subsection{Block diagonalisation argument}\label{sec:block}
  \begin{figure}[h]
  \center \includegraphics[scale=.4]{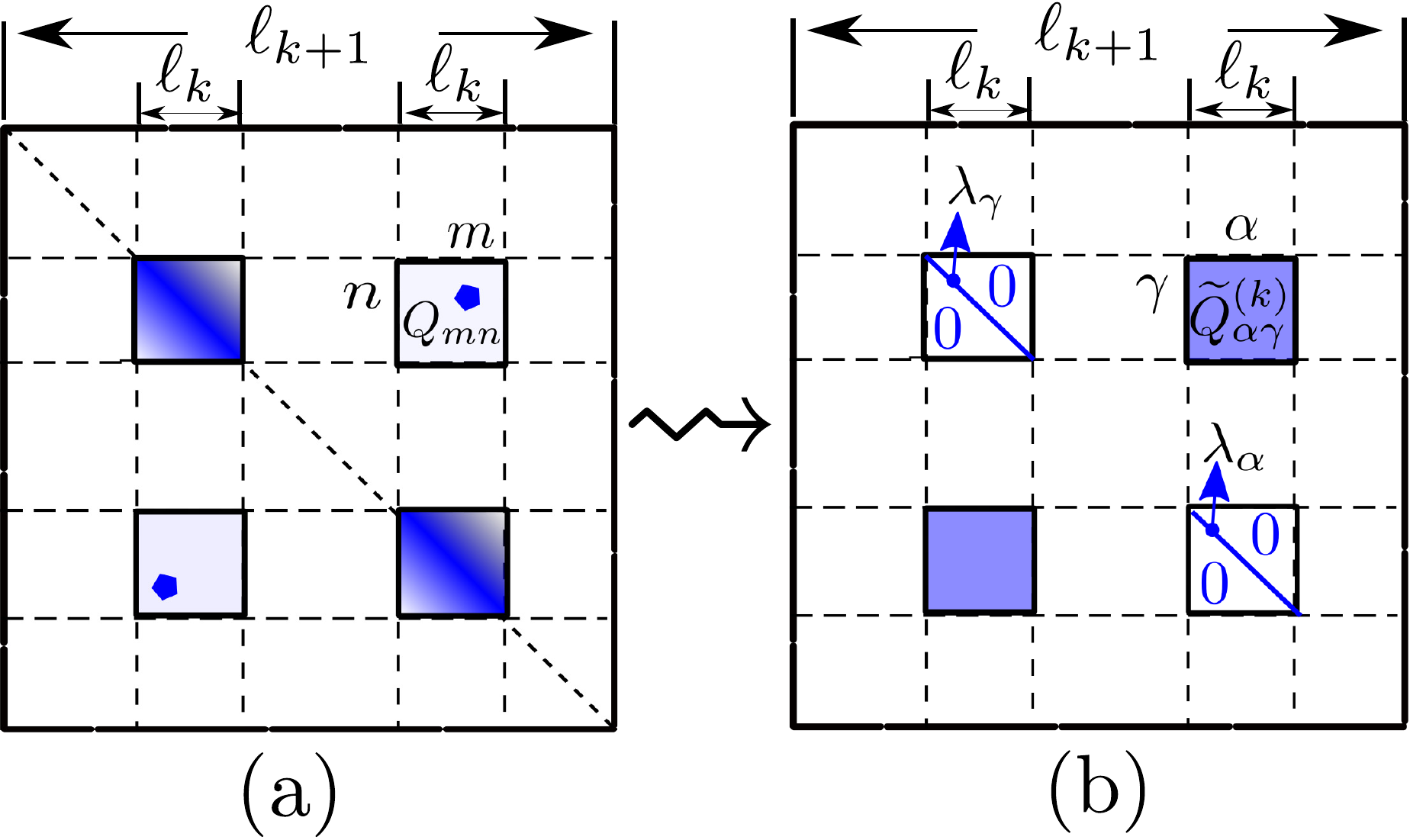}
   \caption{An illustration of the block diagonalisation procedure. (a) depicts a diagonal block of size $\ell_{k+1} \times \ell_{k+1}$, which is divided in into block--matrices of size $\ell_k \times \ell_k$. We pick two blocks of sites labelled by $m$ and $n$ respectively. The diagonal $\ell_k$--blocks are banded matrices (represented as squares with gradient--colour). The choice of the scales (Eq. \eqref{eq:lk}) ensures that the off--diagonal block has a few large elements $Q_{mn} \sim O(1)$ (represented as a dot) while the typical elements are negligibly small. (b) depicts the matrix $ \widetilde Q^{(k)}_{\alpha \gamma}$ (Eq. \eqref{eq:Qtransfred}) after diagonalizing the $\ell_k$--diagonal blocks, so the latter become diagonal. The unitary transformations smear out the quasi--sparseness of $Q_{mn}$ far from the diagonal, and the transformed elements $\widetilde Q^{(k)}_{\alpha \gamma}$ are no longer broadly distributed. }\label{fig:block}
  \end{figure}
The block--diagonalisation renormalisation group procedure considered here is inspired by the statistical models that we study in section \ref{sec:decay}. 
Referring to Figure \ref{fig:block} for an illustration, the basic idea is to take a sequence of lengths $\ell_0 \ll \ell_1 \ll \dots \ll \ell_k \ll \ell_{k+1} \ll \dots $, which will be determined later [Eq. \eqref{eq:lk}]. At step $k$, we divide the matrix $Q$ into blocks of size $\ell_k \times \ell_k$ and let 
$\vert \phi_{\alpha}^{(k)} \rangle$ be the eigenstates of the matrix of diagonal blocks. In that basis, the  matrix elements write as:
\begin{equation}\label{eq:Qtransfred}
\widetilde Q^{(k)}_{\alpha \gamma} =
\sum_{(n,m)} \langle\phi_\alpha^{(k)} \vert n\rangle  Q_{nm} \langle  m \vert   \phi_{\gamma}^{(k)}\rangle \,,
  \end{equation}
 where the sum is over all pairs $(n,m)$, with $n$ ($m$) in the block of $\alpha$  ($\gamma$, respectively). 
 
 The length scale $\ell_{k+1}$ is chosen as a function of $\ell_k$ to ensure that in each off--diagonal block with distance $\leq \ell_{k+1}$ from the diagonal, there is at least one black swan $\abs{Q_{mn}} \sim 1$. By Eq. \eqref{eq:Qmndef0}, this amounts to requiring 
   \begin{equation} \left[\beta \ell_{k+1}^{\mu+1} \right]^{-1} \ell_k^2  = 1 \,,\label{eq:Misg} \end{equation} which gives a sequence of length scales
  \begin{equation}
  \ln (\ell_k / \xi) = \left(\frac{2}{1+\mu}\right)^{k} \ln (\ell_0 / \xi), \quad \xi = \beta^{\frac{1}{1-\mu}} \,,  \label{eq:lk}
  \end{equation} 
 where $\ell_0$ is the initial block size. In order to have an increasing series $\ell_k$, we need $\ell_0 > \xi$ (and $\mu < 1$). Two cases should be distinguished: For weak disorder, $\beta \ll 1$, the initial block size can be small, and its matrix elements of the diagonal blocks are of order unity [by Eq. \eqref{eq:Qmndef0}], so the iteration starts with $\vert \phi_{\alpha}^{(0)} \rangle$ which are extended. On the other hand, for strong disorder, $\beta \gg 1$, $\ell_0$ is large and the diagonal blocks are almost diagonal matrices with localised eigenstates at step $0$.

 In the following, we show that each of the starting situations is preserved under iteration. A crucial step of our argument is the study of the statistical property of the matrix element Eq. \eqref{eq:Qtransfred}, which is a sum of $M = \ell_k^2$ terms, that can be re--written as:
   \begin{align}
  & S[c_i] = \sum_{i=1}^{M} Q_i c_i     \label{eq:defQtilde} 
    \end{align}
   where $ \left\{ c_1, \dots, c_M \right\} =
   \left\lbrace\langle\phi_\alpha^{(k)} \vert n\rangle  \langle  m \vert   \phi_{\gamma}^{(k)}\rangle \right\rbrace $, and $ Q_i $ are the $M$ matrix elements $Q_{mn}$ in Eq. \eqref{eq:Qtransfred}.  Note that the distribution of $Q_{mn}$ depends on $\abs{m-n}$. For simplicity, we approximate all the distances by the maximal value  $\abs{m-n} = \ell_{k+1}$, so that $Q_i$ have identical distribution. Moreover observe that the coefficients $c_i$ depend on the diagonal blocks, so are independent from the $Q_i$ in an off--diagonal block, see Fig \ref{fig:block}. As a helpful warm--up, in Appendix \ref{sec:selfave}, we carry out such a study for the toy--model case in which $c_i = 1$.

   \subsubsection{Extended case}
   Let us assume that at step $k$, the block eigenstates $ \phi_\alpha^{(k)}$ in Eq. \eqref{eq:defQtilde} are extended. Then, $\abs{\langle n \vert \phi_\alpha^{(k)}\rangle}  \sim  \sqrt{1/\ell_{k}}$ independently of $n$. For simplicity, we assume that $c_1, \dots, c_M$ in Eq. \eqref{eq:defQtilde} as identically distributed independent random variables, which have the same distribution as $v / \ell_k$, where $v$ is some random variable with pdf $P(v)$. Then, using Eq. \eqref{eq:broaddef} and then Eq. \eqref{eq:Misg}, we have (we denote $g_{k+1} = \beta \ell_{k+1}^{\mu+1} = M$)
     \begin{align*}
     &\overline{\exp(-t S[c_i])} = \left(\overline{\exp(-v t Q / \ell_k )}\right)^M \nonumber \\
      = 
     & \left( 1 - g_{k+1}^{-1} \int  f(tv / \ell_k) P(v) \dif v  + O(g_{k+1}^{-2}) \right)^M \nonumber \\ 
   &  \stackrel{M\to\infty}{\longrightarrow} \exp(- f_1(t / \ell_k)) \,;\, f_1(s) :=  \int f(s v)  P(v)\dif v  \,.  
     \end{align*}
   This means that $S[c_i] = V / \ell_k$ where $V$ is no longer broadly distributed when $g_{k+1} \to \infty$: in particular, its characteristic function is $\overline{\exp(-t V)} = \exp(-f_1(t))$. Therefore, for any $\ell_{k+1}$--block diagonal matrix, its matrix elements have typical magnitude $\widetilde Q^{(k)}_{\alpha\gamma} \sim 1/\ell_k$, so $\widetilde Q^{(k)}$ is no longer quasi--sparse but rather like the PBRM up to size $\ell_{k+1}$. This allows us to employ the resonance argument in section \ref{sec:intro}, and compare the typical matrix element $\abs{\widetilde Q^{(k)}_{\alpha\gamma}} \sim 1 / \ell_k$ to the typical level spacing on scale $k+1$, which is $\delta_{k+1} \propto 1/\ell_{k+1}$, because the eigenvalues are of order unity and there are $\ell_{k+1}$ of them. Since $\ell_k \ll \ell_{k+1}$, we conclude that $\ell_{k+1}$--block eigenstates are still extended. Repeating this procedure \textit{ad infinitum} we conclude the eigenstates are extended when $\beta \ll 1$. 
   
  We stress that the block diagonalisation procedure is essential for showing the existence of extended phase. Indeed the original matrix is composed by typical matrix elements which are exponentially small and few black swans of order unity. It is precisely the diagonalisation at a smaller scale that spreads the matrix element magnitudes evenly, making them similar to the PBRM.
  
  \subsubsection{Localised case}
    Now we turn to the localised case. Again, we consider the step $k$, and try to estimate the sum Eq. \eqref{eq:defQtilde}. We  assume that the eigenstates $\vert \phi_\alpha^{(k)} \rangle$ are localised, with a decay rate $\abs{\langle \phi_\alpha \vert n \rangle} \sim \pm \exp(- a(\abs{n-n_\alpha}))$, where $n_\alpha$ is the localised centre, and $a(\ell)$ is a growing function of $\ell_k$.
    
    Since $\abs{n-n_\alpha}$ ranges from $1$ to $\ell_k$,  the coefficients $c_i$ in Eq. \eqref{eq:defQtilde} have very different magnitudes: $c_i =  O(1)$ for few $i$'s, and $c_i \sim e^{-2 a(\ell_k)}$ for the other typical $i$'s. Since $c_i$ are uncorrelated from $Q_i$, and $\abs{Q_i} \sim O(1)$ also for a few $i$'s, the event $Q_i c_i \sim 1$ happens with vanishing probability $1/M$. So we are left with two (extreme) types of contributions to the sum \eqref{eq:defQtilde}:
  \begin{itemize}
  \item[(i)] From the terms with $c_i \sim 1$; since there are only $O(1)$ such terms, black swans in $Q_{nm}$ cannot occur (except in rare events of probability $\sim 1/M$), so we have $Q_{nm} \sim Q^{\text{typ}}$ and the total contribution of these terms is $ [\mathrm{I}] \sim  Q^{\text{typ}}.$
  \item[(ii)]  From the terms with typical $c_i \sim e^{-2 a(\ell_k)}$; since there are $\sim M$ of them, there will be $O(1)$ black swans $Q_{nm} \sim 1$, so the total contribution is $ [\mathrm{II}]\sim e^{-2 a(\ell_k)}$.
  \end{itemize}
    Now, since our model is long--range, it is safe to assume that the decay rate of the localised states is no faster than exponential $a(\ell) < c \ell$. Then by eqs. \eqref{eq:Qtyp} and \eqref{eq:Misg}, we have
    $$ \ln \abs{[\mathrm{I}]} \leq -c g_{k+1} \propto  -\ell_{k}^2  \ll - \ell_k  < \ln \abs{[\mathrm{II}]} \,. $$
    This indicates that the sum Eq. \eqref{eq:defQtilde} is dominated by the latter case:
     \begin{equation}\label{eq:Qtildelocalised}
     \widetilde Q^{(k)}_{\alpha\gamma} \sim \exp(-2 a(\ell_k)) \,.
     \end{equation}
    
  Now we apply Eq. \eqref{eq:Qtildelocalised} to estimate the decay of the eigenstates of generation $k+1$,  at first order in perturbation theory. The latter gives: 
  \begin{equation}
 \langle m  \vert \phi^{(k+1)}_\alpha \rangle = \sum_{\gamma} \frac{\widetilde Q^{(k)}_{\alpha\gamma}}{\lambda_{\alpha}- \lambda_{\gamma}} \langle m \vert \phi^{(k)}_\gamma \rangle + \dots \label{eq:1order}
  \end{equation}   
   where the site $m$ belongs to the $\ell_{k}$--block whose eigenstates do not contain $ \vert \phi^{(k)}_\alpha \rangle$ but contains the states $ \vert \phi^{(k)}_\gamma \rangle$. We consider again the two extremal contributions to the sum of Eq. \eqref{eq:1order}:
   \begin{itemize}
   \item[(i)] Since the states $\phi^{(k)}_\gamma$ are localised, $\langle m \vert \phi^{(k)}_\gamma \rangle$ is small except for a few $\gamma$'s localised around $m$. For these $\gamma$'s, energy resonance with $\alpha$ occurs with vanishing probability, thus $\abs{\lambda_{\alpha}- \lambda_{\gamma}} \sim O(1)$. So we have contributions to Eq. \eqref{eq:1order} of magnitude $\sim \widetilde Q^{(k)}_{\alpha\gamma}$.
   \item[(ii)]  On the other hand, if we consider all the $\gamma$'s, the energy mismatch can be as small as $\propto 1/ \ell_{k}$ for a few $\gamma$'s, for which the magnitude of $ \langle m \vert \phi^{(k)}_\gamma \rangle \sim e^{ -a(\ell_{k}) }$, so such contributions to Eq. \eqref{eq:1order} have magnitude $\sim \widetilde  Q^{(k)}_{\alpha\gamma} \ell_{k} e^{-a(\ell_{k})}$. 
   \end{itemize}
  Now we make the assumption that $a(\ell) \gg \ln \ell$, \textit{i.e.}, the eigenstates decay faster than algebraically, the first kind of contribution dominates, giving 
     \begin{align*}
     & e^{-a(\ell_{k+1})} = \langle m  \vert \phi^{(k+1)}_\alpha \rangle  = \widetilde Q^{(k)}_{\alpha\gamma} \sim \exp(-2 a(\ell_k))  \\
     \Rightarrow & a(\ell_{k+1}) = 2 a(\ell_k) \,. 
     \end{align*}
  The solution to this recursion relation is:
   \begin{align}   
  & a(\ell) \propto \ln^{\kappa(\mu)} (\ell /\xi) \,,\, \nonumber \\
  &  \kappa(\mu) = \frac{\ln 2}{\ln \frac{2}{1 + \mu}} > 1 \,, \label{eq:kappamu}
   \end{align} 
   justifying the assumption $a(\ell) \gg \ln \ell$: Eq. \eqref{eq:kappamu} is self--consistent. Remark that by Eq. \eqref{eq:Qtildelocalised}, $\widetilde Q^{(k)}_{\alpha\gamma} \ll \ell^{-y}$ for any $y$: the transformed hopping elements decay faster than any power--law, thus making the PRBM--like resonance impossible.
   
   To recapitulate, we have shown that for $\beta$ large enough, the eigenstate is localised with a peculiar decay 
   \begin{equation}
   \abs{\langle n \vert \phi \rangle} \sim \exp\left( - C \ln \abs{n - n_{\max}}^{\kappa(\mu)}\right) \,,  \label{eq:decayLT}
   \end{equation}
   where $C$ is some constant. In section \ref{sec:decay}, we shall return to the decay of localised states, and extend the above result to $\mu > 1$.
   
\textit{Discussion.} Summarising the two cases, we have shown the existence of localisation transition and the peculiar decay rate of the localised states. In renormalisation--group terms, we have shown that the $\beta \to \infty$ and $\beta \to 0$ limits are attractive fixed points.  
 
It is more subtle to argue for the existence or the absence of \textit{mobility edges}, \textit{i.e.}, localisation transition by varying the $\lambda$ (not $\beta$). Indeed, in the above arguments, the dependence on $\lambda$ is implicitly present when we make comparisons to the level spacing. The latter depends on the DoS, which in turn depends on $\lambda$. Therefore, it is reasonable to argue that the critical disorder strength $\beta_c$, which depends on the properties of the matrix ensemble close to the diagonal, has also a non--trivial $\lambda$ dependence: $\beta_c = \beta_c(\lambda) \neq \text{constant}$. As a consequence, for at least some values of $\beta$, the matrix ensemble with disorder parameter $\beta$ has mobility edges at some $\lambda_c$. By this argument, we expect that the existence of mobility edges is \textit{generic} in the broadly distributed class. To support our claim of mobility edges, and further characterise the phase diagrams, we shall study numerically the two particular models in the next section.

 \subsection{Numerical study}\label{sec:num}
 The most common probe of the localisation of eigenstates is the inverse participation ratio. Recall that for a normalised state $\phi$, it is defined as
    \begin{equation}
    P_2 =  \sum_{n=1}^N \abs{\langle \phi\vert n \rangle}^{4} \,.
    \end{equation}  
  The asymptotic behaviours (as $N \to \infty$) of $P_2$ the extended and localised phase are
  \begin{equation}\label{eq:Pqphases}
   P_2 \sim \begin{dcases}
    N^{0} & \text{localised phase} \,, \\ N^{-1}  & \text{extended phase} \,.
   \end{dcases}
  \end{equation}
 At the localisation transition, $P_2 \sim N^{-\tau_2}$ for some non-linear exponent function $\tau_q$ characterising the multi--fractal properties of the critical eigenstate. 
 
\begin{figure*}
   \center  \includegraphics[width=.85\textwidth]{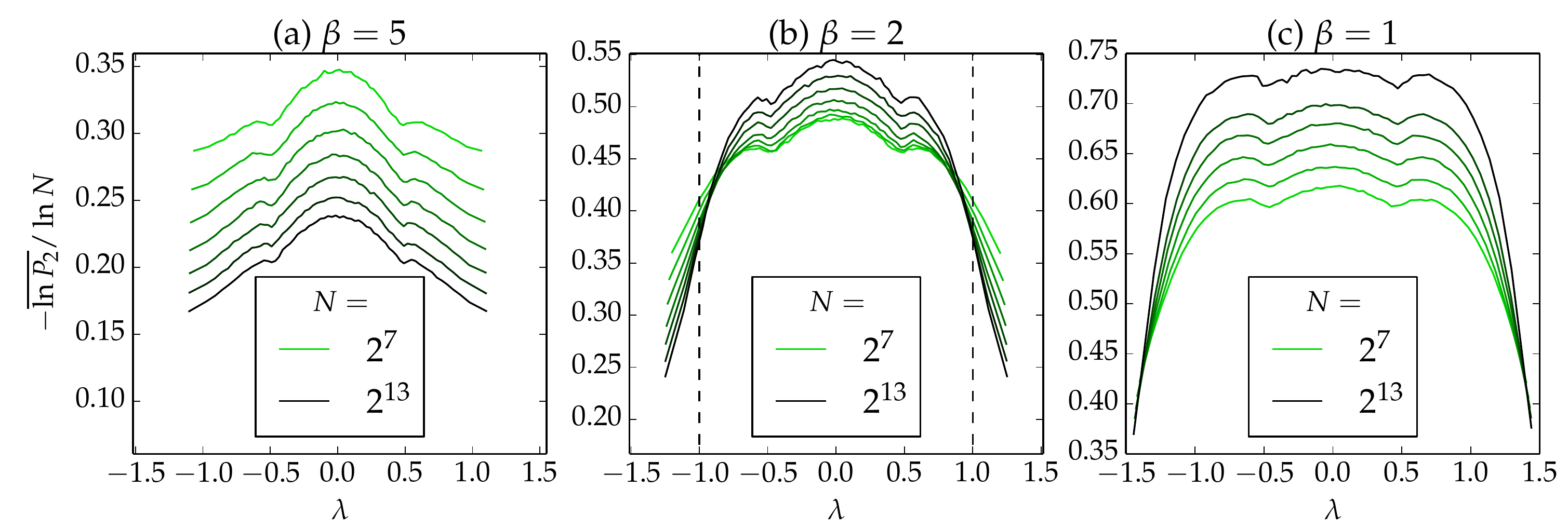}
   \caption{The IPR of the eigenstates of the RSM, plotted for parameters $\mu = 0.5$, $\beta = 1, 2$ and $5$, for matrix sizes $N = 2^7, \dots, 2^{13}.$ The numerical protocol is identical to that of Figure \ref{fig:IPR}. We recall the for ideally extended (localised) states, the $y$ value is $1$ ($0$, respectively). In panel (b), the estimated position of mobility edges $\lambda_c = \pm 1.0(2)$ is also indicated. In all plots, lighter colours represent smaller system sizes.} \label{fig:sparse}
   \end{figure*}

\begin{figure}
     \center  \includegraphics[width=.49\columnwidth,valign=t]{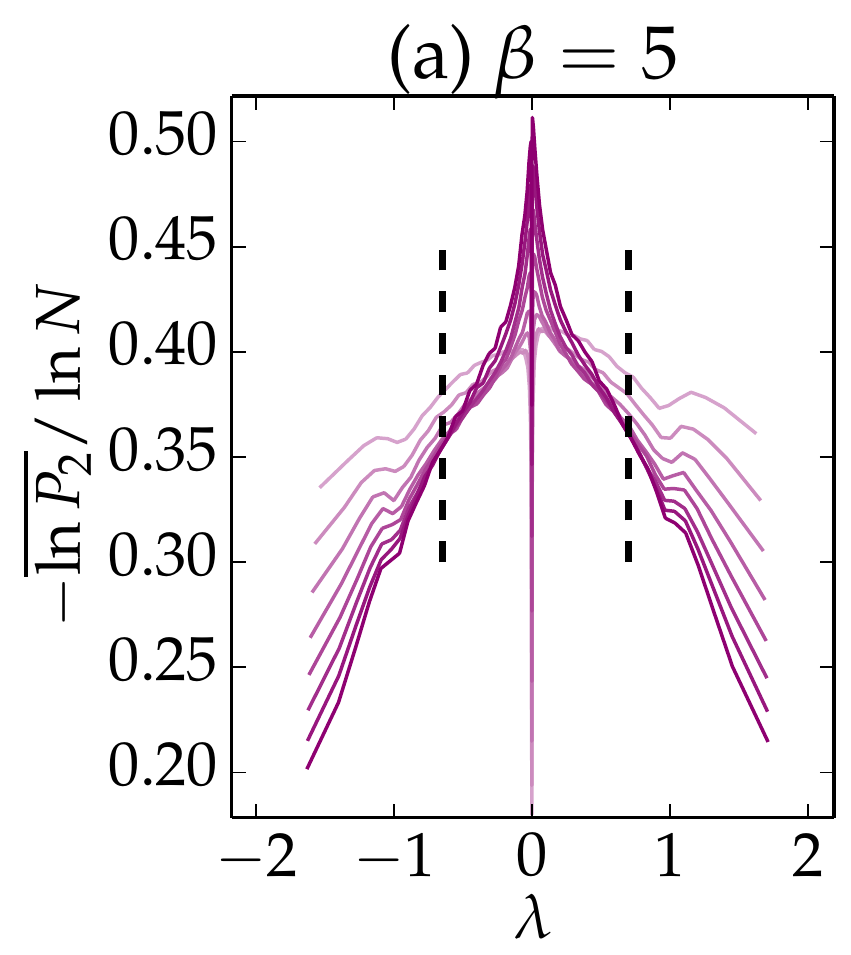} 
      \includegraphics[width=.49\columnwidth,valign=t]{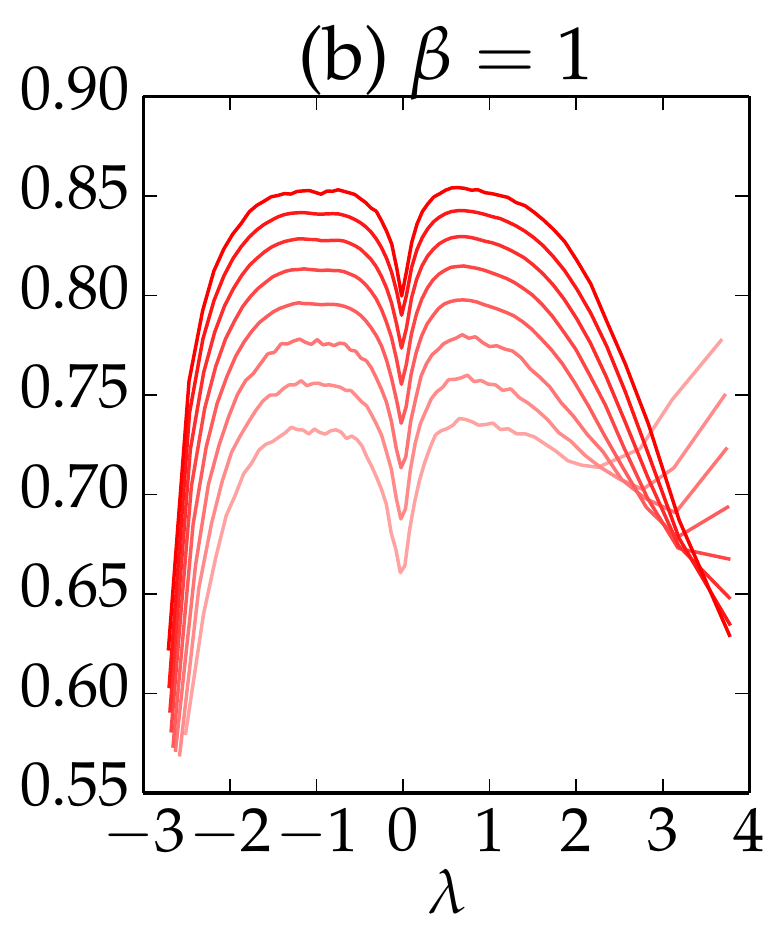}
     \caption{ Measure of the IPR of  BBRM eigenstates as function of eigenvalue $\lambda$, at $\mu = 0.5$ and with various sis of matrices, $N = 2^7, \dots, 2^{14}$ (darker colours indicate larger systems). According to Eq. \eqref{eq:Pqphases}, extended (localised) states would have $y$ coordinate $-\overline{\ln P_2}/\ln N \to 1$ ($\to 0$, respectively) when $N \to \infty$. (a) $\beta = 5$. The dashed lines indicate the mobility edges, estimated at $\lambda_c \approx -0.65(5)$ and $\lambda_c \approx 0.70(5)$. (b): The same measure for $\beta = 1$. As $N$ increases (darker colour), the eigenstates with $\lambda \in (-3,3)$ becomes more and more de-localised. } \label{fig:IPR}
    \end{figure}

To measure numerically the IPR, we generated samples of the BBRM and randomised sparse matrices at $\mu = 0.5$, for different values of $\beta$ and different system sizes, and exactly diagonalised them using standard routines. The results lead to the qualitative phase diagrams of Figure \ref{fig:sparsephase}. A selection of data are shown in Figure \ref{fig:sparse} (RSM) and \ref{fig:IPR} (BBRM), and we discuss the salient features below.   

\subsubsection{Mobility edges}
For both BBRM and RSM, we found that for some value of $\beta$ (we show data for $\beta = 5$ for BBRM and $\beta=2$ for RSM), the spectrum is separated into de--localised state in the middle and localised states near the edges. The estimated positions of the mobility edges are indicated in Figure \ref{fig:sparse} (b) and \ref{fig:IPR} (a). In both models, we observe a pronounced finite size effect: near the transition, and at the extended side, the system would seem localised if only small systems were considered. We can understand this effect in light of the argument in section \ref{sec:block}: the hierarchy of scales grows very fast, Eq. \eqref{eq:lk}, and the delocalisation takes place thanks to the presence of the \textit{rare} elements far from the diagonal: both mechanisms are probably missed in small systems. In this respect the localisation transition of the broadly distributed class is akin to that in Lévy matrix, as carefully characterised recently in \cite{tarquini16levy}. 

\subsubsection{Lower--critical disorder}
In both BBRM and the randomised sparse matrices (with $\mu=0.5$), we observe that when $\beta=1$, all the states tend to be extended, with the possible exception of those at the edges of the spectrum. Thus, we suspect that there is a lower--critical disorder $\beta_{-}$, below which the mobility edges disappear. 

If this is indeed the case, the phase diagram of the broadly distributed class at small disorder will be different from that of the finite--dimensional Anderson model (for which $\beta_- = 0$), and resembles that of the Anderson model on the Bethe lattice \cite{aizenman11tree,aizenman11treemob}, which is known to have $\beta_->0$. 

However, in the absence of analytical arguments, the numerical data at hand do not allow a definite conclusion concerning whether $\beta_- > 0$. We leave this open question for future study. 

\subsubsection{Upper--critical disorder}
More can be said about the existence of an upper--critical disorder $\beta_c < \infty$, beyond which all the states become localised. Here, BBRM and RSM  are qualitatively different: For RSM, the upper--critical disorder certainly exists: $\beta_c < 5$, since for $\beta=5$, all the states are already localised, see Figure \ref{fig:sparse} (c). 

For the BBRM, extended states are observed near $\lambda = 0$ for all the values of $\beta$ we considered (up to $\beta=10$, data not shown): mobility edges never disappear. Indeed, in appendix \ref{sec:LRP-ME}, we argue that $\beta_c = \infty$ for BBRM. Understanding completely the argument requires the mapping to statistical models, to be discussed in section \ref{sec:decay}. Yet, roughly speaking, $\beta_c = \infty$ is due to the absence of diagonal disorder and the divergence of the DoS of BBRM. In general, we should however be careful on relating the absence of diagonal disorder to $\beta_c = \infty$. One difficulty is the nature of the singularity at $\lambda = 0$. Indeed, for the variants of randomised sparse matrix whose near--diagonal elements are also sparse, we observe numerically that the DoS has an even stronger divergence at $\lambda = 0$, resembling a $\delta$ peak. Such a singularity makes the notion of mobility edge problematic: \textit{e.g.} there could be a transition inside the finite portion of spectrum concentrated at $\lambda = 0$, and which cannot be described by any $\abs{\lambda_c} > 0$.

\subsubsection{Level statistics}
Another numerical method to characterise the localisation transition is the level statistics of eigenvalues. Here, we shall consider the BBRM and the \textit{gap ratio} observable proposed in \cite{oganesyan07huse}. Denoting $\lambda_1 \leq \dots  \leq \lambda_N$ the ordered eigenvalues of a matrix, and $\delta_i = \lambda_{i+1} - \lambda_i$ the level spacings (gaps), we consider the following ratio between successive gaps 
\begin{equation} r_i = \min(\delta_i,\delta_{i+1})/\max(\delta_i,\delta_{i+1}) \,. \end{equation}
The advantage of this observable is that it does not depend on the DoS. Its mean value is universal in localised and extended phases~\cite{atas13ratio}:
\begin{equation}
\overline{r} = \begin{dcases}
 r_P  = 2 \ln - 1 \approx 0.39   & \text{localised phase} \,, \\
 r_{GOE} \approx 4 - 2\sqrt3 \approx 0.53 \,,  & \text{extended phase}.
\end{dcases}
\end{equation} 
The localised value $r_P$ comes from Poisson level statistics, while in the extended phase, $r_{GOE}$ is the value of the GOE ensemble; its approximate and numerical values were studied in \cite{atas13ratio}. It is then convenient to define the rescaled gap ratio
 \begin{equation} \chi \defeq \frac{\overline{r} - r_{P}}{r_{GOE} - r_P} \Rightarrow \chi \rightarrow \begin{dcases} 0 &  \text{localised/Poisson}, \\ 1 & \text{extended/GOE}. \end{dcases} 
 \label{eq:defchi}\end{equation} 
 
 \begin{figure}
 	\center
 	\includegraphics[width=.47\columnwidth,valign=t]{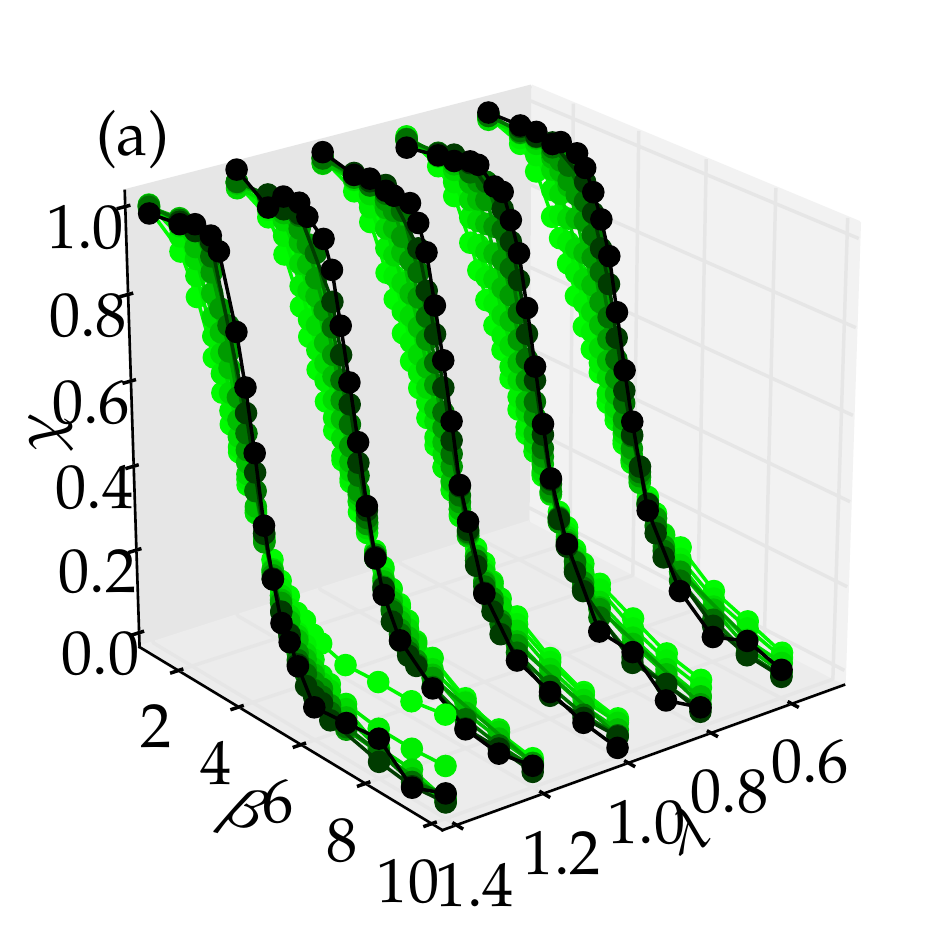}
    \includegraphics[width=.51\columnwidth,valign=t]{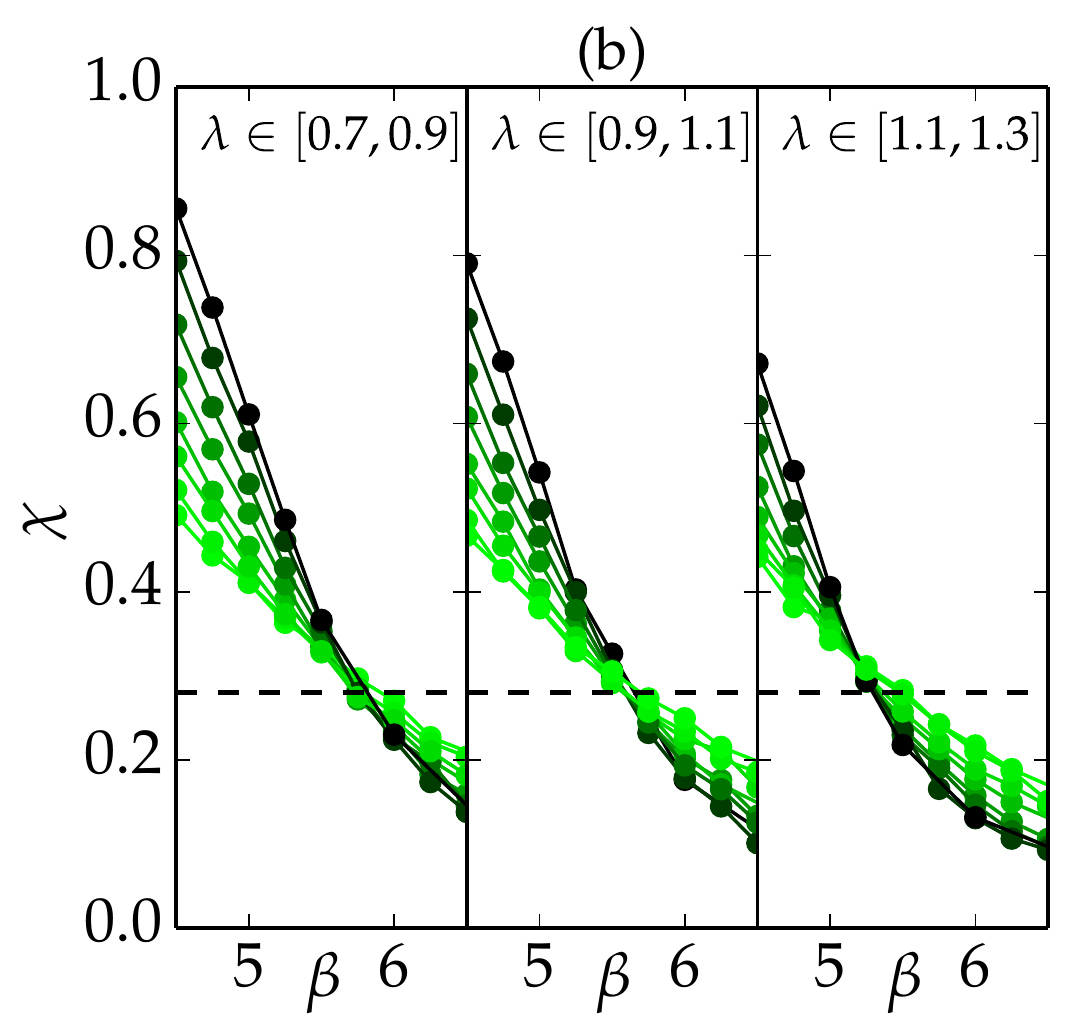}
 	\caption{(a) Numerical measure of ratio $\chi$, Eq. (\ref{eq:defchi}) for the BBRM model with $\mu = 0.5$, $\beta \in (1,10)$ and of sis $N = 2^7$ (light color) to $2^{14}$ (dark color). Eigenvalues in $(0.5, 1.5)$ are binned into $5$ bins of equal width. More than $10^5$ different gaps are averaged over for each data point.  (b) A zoom-in to the critical regime. We locate the critical value of the rescaled gap ratio observable $\chi_c \approx .28(2)$.}\label{fig:transition}
 \end{figure}
We measured this quantity for BBRM with $\mu = 0.5$ for several values of $\beta \in [1,10]$, and for a few windows of $\lambda$ of width $0.2$. As shown in Figure \ref{fig:transition} (a), $\chi$ goes from the GOE (extended) value to the Poisson (localised) value when $\beta$ increases from $1$ to $10$, and the change becomes sharper as the system size increases. In Figure \ref{fig:transition} (b), we look more closely at $\beta \sim 5.5$ and three windows of eigenvalues to examine the critical region. We observe that the critical value of the rescaled gap ratio $\chi_c \approx 0.28(2)$ is independent of $\lambda$; this indicates that for a given $\mu \in (0,1)$, there is one unique critical point of localisation transition. 

On the other hand, there is a quantitative discrepancy between the mobility edge positions estimated by the IPR and the level statistics. In fact, we observe that $\abs{\lambda_c^{\text{IPR}}} < \abs{\lambda_c^{\text{ratio}}}$, \textit{i.e.}, there is a critical region which seems to be localised according to IPR but extended according to level statistics criterion. Such a discrepancy has been observed in other matrix models, like the Lévy random matrix \cite{cizeau1994levy,tarquini16levy,monthus16levy} and the Anderson model on the Bethe lattice \cite{biroli2012difference}, and was interpreted as a signature of a mixed phase. In particular, for the Bethe lattice, it was proposed that this phase is \textit{critical} and characterised by multifractal eigenstates  \cite{deluca14bethe,altshuler2016rrg,altshuler16rrg1}. However, the existence of the critical phase is highly controversial \cite{mirlin16rrg,metz2017level}, and the discrepancy described above has also been explained in terms of a large finite size effect \cite{garcia2016scaling,tarquini16levy}. 
In light of these lessons, we shall refrain from advancing any statement concerning critical phase before further study. 

 \section{Localised state decay and relation to statistical models}\label{sec:decay}
 In this section, we discuss the decay of localised states of the broadly distributed class using mappings between them with two statistical models: the epidemic growth model studied in \cite{hallatschek2014acceleration,chatterjee2016multiple} and the long--range percolation \cite{newman1986lrp}. Section \ref{sec:fisher} considers the specific case of BBRM and its exact mapping to the epidemic model. Section \ref{sec:LRP} relate the broadly distributed class in general to the long--range percolation model. Finally section \ref{sec:dimension} interprets the results in terms of effective dimension.
 
 \subsection{Epidemic growth and BBRM}\label{sec:fisher}
 \begin{figure}
 \center 
 \includegraphics[scale=.8]{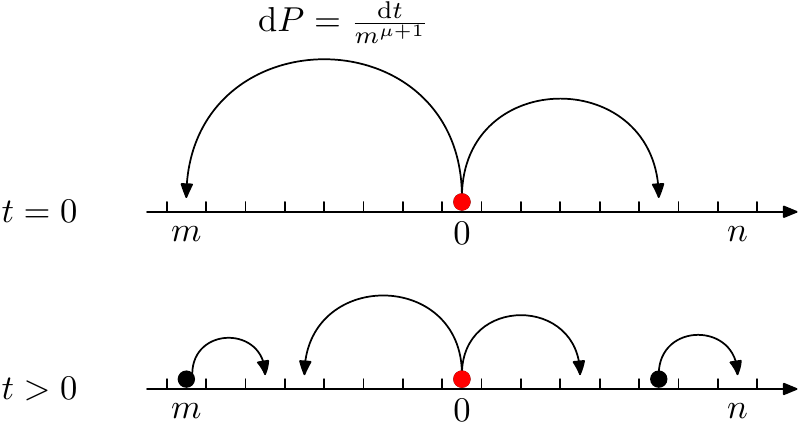}
 \caption{An illustration of the dynamics of the epidemics dynamics model of \cite{hallatschek2014acceleration} in $d=1$. Initially ($t = 0$), the only occupied site at $n = 0$ infects other sites at a rate that decays as Eq. \eqref{eq:infectionrate}.  Later, $t > 0$, infected sites go on to infect remaining sites.}\label{fig:fishermodel}
 \end{figure}
 Fisher and Hallatschek studied in \cite{hallatschek2014acceleration} an epidemic growth model with long--range dispersion. We recall its definition on a 1d lattice. Every site can be either empty or occupied (infected). The dynamics is as follows, see Figure \ref{fig:fishermodel} for an illustration: 
 \begin{itemize}
 \item Initially, at $t = 0$, only the site $0$ is occupied, and all the others are empty.  
 \item During each infinitesimal time interval $\dif t$, and for any pair of sites $m$ and $n$, such that $m$ is occupied and $n$ is empty, $n$ becomes occupied with probability rate
 \begin{equation}\label{eq:infectionrate}
 \dif P_{mn} = \abs{m-n}^{-\mu - 1} \dif t  \,.
 \end{equation} 
 That is, occupied sites ``infect'' all other sites with a rate that decays algebraically as function of the distance. All these events are uncorrelated.
 \item Once a site is occupied, it remains so forever. 
 \end{itemize}
 
\begin{figure*}
\includegraphics[scale=1]{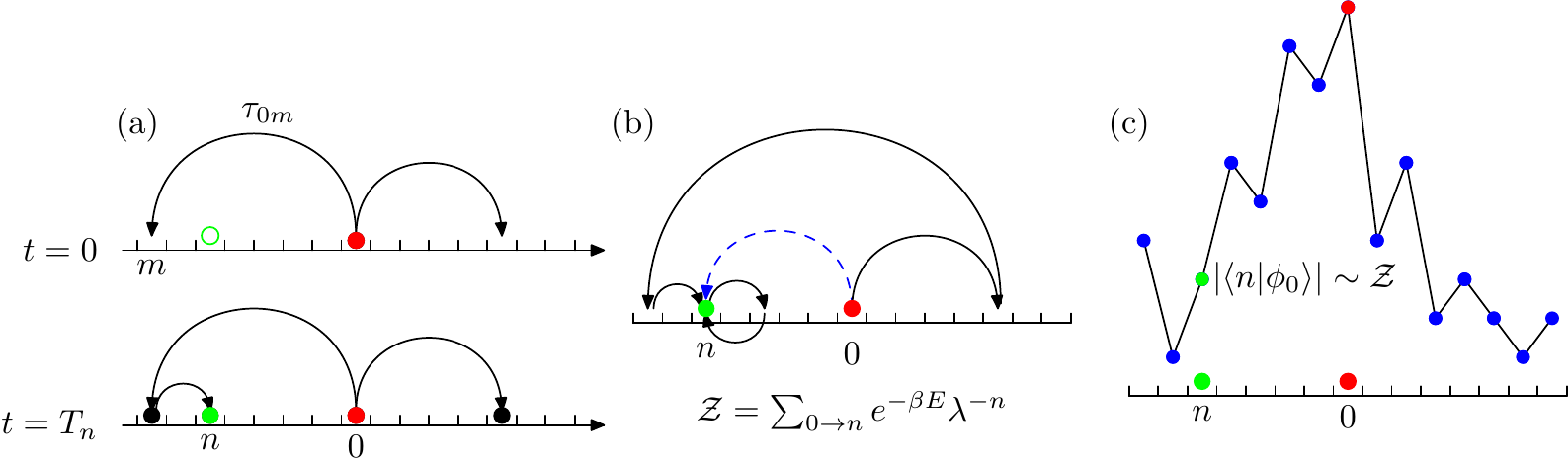}
    	\caption{Illustration of models involved in the mapping described in Sect. \ref{sec:fisher}. (a) In the FPP model, a waiting time $\tau_{mn}$ is assigned to each pair. The first--passage time $T_n$ is a minimum of all paths, given by Eq. \eqref{eq:Esupp}. (b) In the polymer model, which is a finite temperature extension of the FPP model, we sum over all paths connecting two points $0$ and $n$, ranging from the direct path (blue, dashed) and detoured paths with loops, for example, the one in black. (c): The amplitude at site $n$ of (strong-disorder) BBRM eigenstates localised around $0$ is in turn related to the polymer partition function by \eqref{eq:roTsccarsup}.} \label{fig:trio}
    \end{figure*}

 A central question of this model is how fast the epidemic colony grows, and can be addressed by considering the \textit{first--passage time} $T_n$, \textit{i.e.}, the first moment where the site $n$ is infected. An equivalent definition of $T_n$ is as follows: let the \textit{waiting times} $\tau_{mn}$ be  independent exponential random variables with mean value $\abs{m-n}^{\mu + 1}$ [as in Eq. \eqref{eq:Qdef2}], then $T_n$ is the following minimum over all the sequences $\mathfrak{p}=(0 = m_0, m_1,\dots, m_s = n)$ connecting $0$ and $n$ (its length $s$ is unconstrained): 
 \begin{equation}
 T_n = \min_{\mathfrak{p}} T[\mathfrak{p}] \,,\, T[\mathfrak{p}] := \sum_{i=1}^s \tau_{m_i, m_{i-1}} \,. \label{eq:Esupp}
 \end{equation}
 Such expression is known in general to relate exactly  growth models and \textit{first--passage percolation} (FPP) models. The \textit{long--range }FPP obtained here, illustrated in Figure \ref{fig:trio} (a), has been studied independently by Chattajee and Dey \cite{chatterjee2016multiple}, whose rigorous results agree with those of Fisher and Hallatschek \cite{hallatschek2014acceleration}.
 
 Now we describe the mapping between the long--range FPP and the BBRM model, \textit{via} an intermediate long--range ``polymer'' in random media model, see Figure \ref{fig:trio} for an illustration. The polymer model is a finite--temperature extension of the FPP, defined by the following grand--canonical partition function:
 \begin{align}
   \mathcal{Z} & = \sum_{s=0}^{\infty} \sum_{\substack{\mathfrak{p}:(m_0, \dots, m_s) \\ m_0 = 0, m_s = n}}
   \exp(- \beta T[\mathfrak{p}] + s \beta \tau ) \nonumber \\
   & = \sum_{s=0}^{\infty} \sum_{m_1} \dots \sum_{m_{s-1}} \exp\left(- \beta \sum_{i=1}^s (\tau_{m_i m_{i-1}} - \tau) \right)    \,. \label{eq:ZofBBRM}
 \end{align}
 Here, the energy associated to a polymer $\mathfrak{p}$ is $E = T[\mathfrak{p}]$; $\beta$ is the inverse temperature, and $\tau$ is the chemical potential coupled to the length $s$. 

 It follows from the definition of the polymer model and Eq. \eqref{eq:Esupp} that the first--passage time is the free energy in the zero--temperature, zero--chemical--potential limit: 
 \begin{equation} T_n =  \left[ - \beta^{-1} \ln \mathcal{Z}  \right]_{\beta \to \infty, \tau\to 0 } \,. \label{eq:zeroTlimit}\end{equation} On the other hand, recall from Eq. \eqref{eq:Qdef2} the BBRM matrix elements $Q_{mn} = e^{-\beta \tau_{mn}}$ are precisely the Boltzmann weight associated to a monomer $m\to n$. Using this fact, and the definition of matrix multiplication, one can rewrite $\mathcal{Z}$ as a resolvant:
  \begin{align}
 \mathcal{Z}  &=  \sum_{s=0}^{\infty} \sum_{m_1} \dots \sum_{m_{s-1}} \prod_{i=1}^s  \left(Q_{m_i m_{i-1}} \lambda^{-1} \right) \nonumber\\ 
  & = \langle n \vert ( 1 - \hat{Q} / \lambda )^{-1} \vert 0 \rangle \,,\, \lambda = e^{-\beta\tau} \,. \label{eq:ZofBBRM2}
  \end{align} 
  Therefore, for any $\lambda \neq 0$ fixed, as $\beta\to \infty$, $\tau \to 0$, and we have by Eq. \eqref{eq:zeroTlimit}
  \begin{equation}
  \langle n \vert ( 1 - \hat{Q} / \lambda )^{-1} \vert 0 \rangle \stackrel{\beta \gg 1}\approx - \beta T_n \,,\, \lambda \neq 0 \,. \label{eq:logresolvant}
  \end{equation}
   The decay of the left hand side (as $\abs{n} \to \infty$) is related that of eigenstates in a quite standard way, which we briefly recall in Appendix \ref{sec:resolvant}:
     \begin{equation}
     - \ln \abs{ \langle n\vert\phi_0 \rangle} \stackrel{\beta \gg 1}\approx \beta  T_{n}  \,,\, \label{eq:roTsccarsup}
     \end{equation} 
 In particular, if $T_{n}$ does not increase with $\abs{n}$, the state $\vert \phi_0 \rangle$ in fact extended. As we have noted, the above equation is valid for $\lambda \neq 0$. More precisely, the requirement is that $\abs{\tau} = \abs{\beta^{-1} \ln \lambda} \ll 1$. So for a given large $\beta$, Eq. \eqref{eq:roTsccarsup} covers the whole spectrum except an exponentially small interval around $0$. 
   
\begin{figure}
 \center \includegraphics[width=.8\columnwidth]{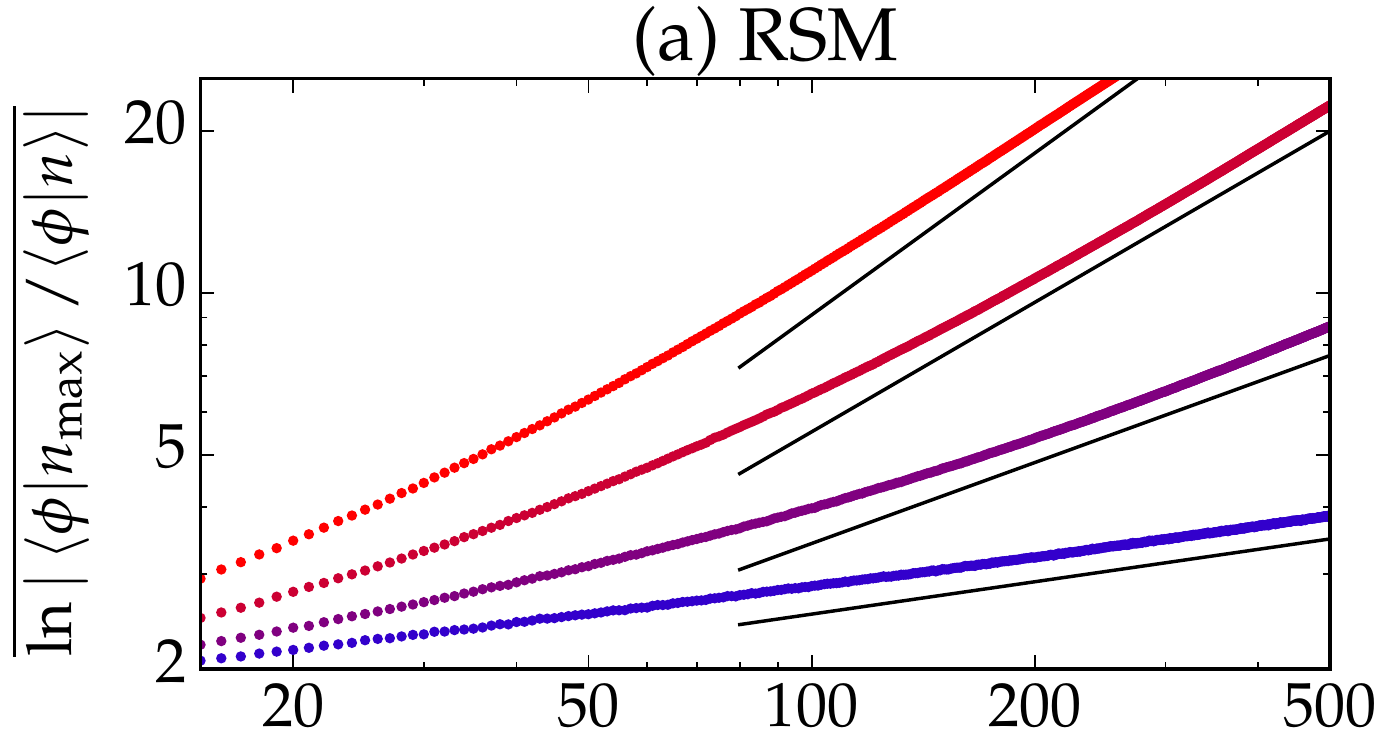}
   \includegraphics[width=.8\columnwidth]{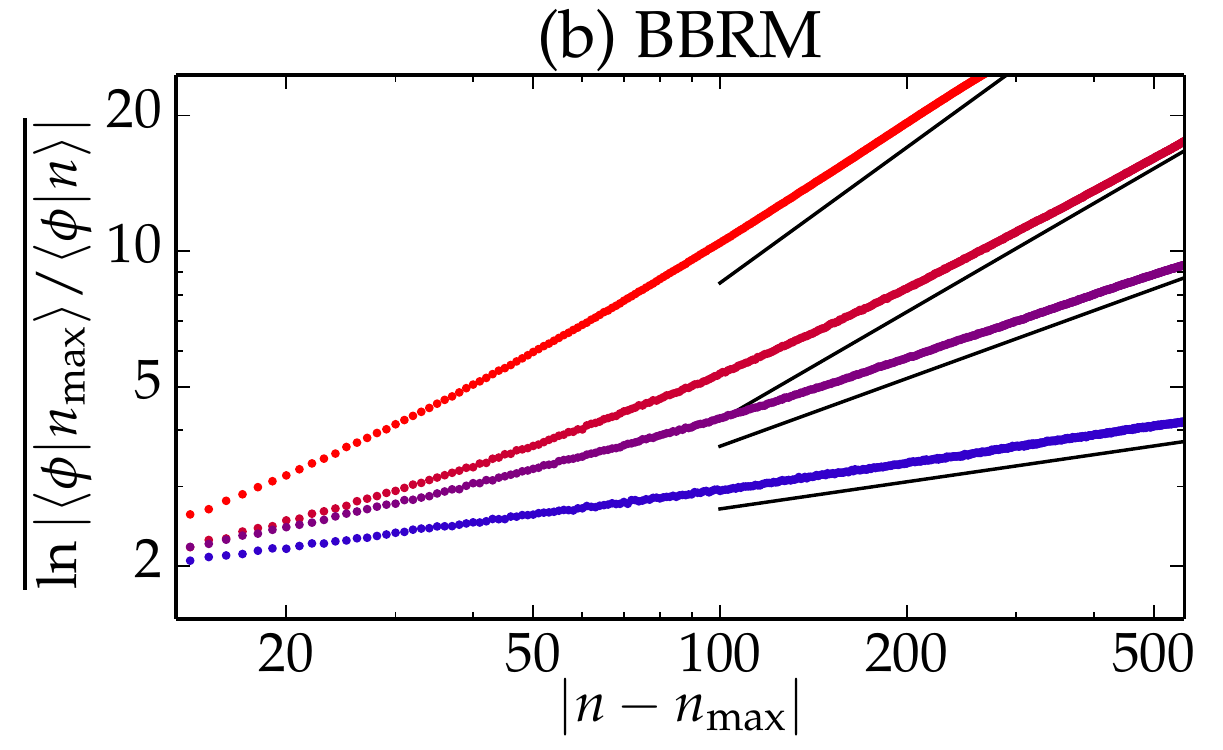}
   \caption{Numerical measure of the decay of the RSM model (a) and the BBRM (b) model in the $\mu > 1$ regime. The dots are the numerical data, and the thin lines represent the prediction Eq. \eqref{eq:stretch}. for all measures $N = 2^{11}$, and the $N/4$ states in the centre of the spectrum of every realisation are measured. (a) From top to bottom: $\mu = 2.2, 1.8, 1.5, 1.2$. For all data, $\beta = 0.1$, $N = 2^{11}$. (b) From top to bottom: $\mu = 2.5, 1.8, 1.5, 1.2$, $\beta = 0.2, 0.2, 0.3, 0.3$.} \label{fig:decay}
   \end{figure}

Eq. \eqref{eq:roTsccarsup} is the main result of this section, and allows to translate the exact asympotitcs of $T_n$ established in \cite{hallatschek2014acceleration,chatterjee2016multiple} into the decay rates of the localised states in the strong disorder $\beta \gg 1$ regime. The results are summarised in Table \ref{table:summary}. When $\mu < 0$, $T_n$ does not grow with $n$, hence the states are in fact delocalised. In the regime $\mu \in (0,1)$, we confirm the prediction Eq. \eqref{eq:decayLT} obtained in section \ref{sec:block}. 
For $\mu > 1$, we predict that the localised states decay in a (stretched)--exponential fashion: 
\begin{equation} \abs{\langle n \vert \phi \rangle  } \sim  e^{-C \abs{n - n_{\max}}^{\alpha} } \,,\, \alpha = \begin{cases}
\mu - 1  & 1 < \mu < 2 \\ 1 & \mu > 2  
\end{cases}  \label{eq:stretch}
\end{equation} where $n_{\max}$ is the localisation centre of $\phi$. Note that, so far, this prediction is obtained only for the BBRM model in the limit $\beta \gg 1$. Nevertheless, we claim that Eq. \eqref{eq:stretch} holds generally for the broadly distributed class in the $\mu > 1$ regime, for any disorder strength $\beta > 0$; in particular, all the states are localised.  These claims are well supported by the numerical simulations on the BBRM and the random sparse matrix model, see Figure \ref{fig:decay}, and will be better understood in light of the more general mapping between broadly distributed class and the long--range percolation models, which we discuss in section \ref{sec:LRP}.
   
To conclude this section, we remark that the mapping above is a new instance of the well-known interplay between polymer in random media and localisation \cite{nss85hops,muller13magneto,pietracaprina16forward,tarquini16levy}, a highlight of which has been relating the conductance fluctuations in short-range Anderson model to Kardar-Parisi-Zhang (KPZ) \cite{kardar1986dynamic} universality class \cite{medina89anderson,somoza2007universal,somoza15anderson}.    
   
   \begin{table*}
     \center	\begin{tabular}{c|c|c|c}
     		$\mu \in$ & $T_{n} \propto$ & BBRM $\abs{\langle n \vert \phi_0 \rangle}\sim$ &    PBRM~\cite{mirlin96pbrm} $\abs{\langle n \vert \phi_0 \rangle}\sim$  \\ \hline
     		$(-1,0]$ & $ \stackrel{N\rightarrow\infty}\longrightarrow 0$ & extended &  extended \\  
     		$(0,1)$ &  $\left(\ln \abs{n}\right)^{\kappa(\mu)}$  & $ \exp\left(- c \ln^{\kappa(\mu)}\abs{n}\right)$/extended & extended  \\ 
     		$(1,2)$ & $\abs{n}^{\mu - 1}$   & $ \exp\left(- c \abs{n}^{\mu-1)}\right)$ &   $ \abs{n}^{-(\mu+1)/2}$ \\
     		$(2, \infty)$ & $\abs{n}$  &  $\exp\left(- c \abs{n}\right)$  & $ \abs{n}^{-(\mu+1)/2}$\\ \hline
     	\end{tabular}
     	\caption{Summary of  the asymptotic first-passage times obtained in  \cite{hallatschek2014acceleration,chatterjee2016multiple} and their implication on the decay of the eigenstates $\phi_0$ of BBRM localised around $0$. $\kappa(\mu) = \ln 2/\ln \frac{2}{\mu + 1}$ as in Eq. \eqref{eq:kappamu}. The results on  PBRM \cite{mirlin96pbrm} are also shown for comparison. In the $\mu \in (0,1)$ regime, the  BBRM has a localisation transition and extended eigenstates. In the $\mu < 0$ regime,  $T_{n} \rightarrow 0$ in the $L\rightarrow \infty$ limit. By \eqref{eq:roTsccarsup}, this means the eigenstates are extended even in the $\beta \gg 1$ limit. }\label{table:summary}
   \end{table*} 
   
 \subsection{Long--range percolation} \label{sec:LRP}
Recall from Eq. \eqref{eq:Qmndef0} that any broadly distributed banded random matrix $Q$ is quasi--sparse. So we can define a random graph $G$, whose vertices are $1, \dots, N$, and $m$ and $n$ are connected if $Q_{mn} \sim O(1)$, that is, with probability 
\begin{equation} p_{mn} = \beta^{-1} \abs{m-n}^{-\mu-1} \,,\, \abs{m-n} \gg 1 \,. \label{eq:pmn} \end{equation}
Such a random graph defines the long--range percolation model, first studied in \cite{schulman83lrp,newman1986lrp,aizenman1986lrp}.

The long--range percolation model is closely related to the epidemic model in section \ref{sec:fisher}. Indeed, it is known \cite{biskup11diameter,biskup2004} that, the \textit{chemical distance}, defined as the length of the shortest path connecting $0$ and $n$ on $G$, and denoted $d_n$, has the same asymptotic growth as the first passage time $T_n$ of the epidemic model, defined in Eq. \eqref{eq:Esupp} \footnote{Strictly speaking, we need to ensure $G$ is connected, by setting $p_{mn} = 1$ for $\abs{m-n} = 1$.}: 
\begin{equation} d_n \propto T_n \,,\, n \to \infty \,. \label{eq:dnTn} \end{equation}
This result explains why the decay rates in Table \ref{table:summary} apply universally to the broadly distributed class. Indeed, if we neglect the small matrix elements in quasi--sparse models such as BBRM, any broadly distributed matrix $Q$ can be seen as a \textit{short--range} Anderson model \textit{on the random graph $G$}, with random hopping elements of order unity: In this respect, the decay rate  of Eq. \eqref{eq:roTsccarsup} is equivalent to an exponential decay with respect to the graph distance: $$ \ln \abs{\langle n \vert \phi_0 \rangle} \sim e^{-c T_n} \sim e^{- c' d_n } \,,$$
which is naturally expected for a short--range model. 

\begin{figure}
\includegraphics[width=.9\columnwidth]{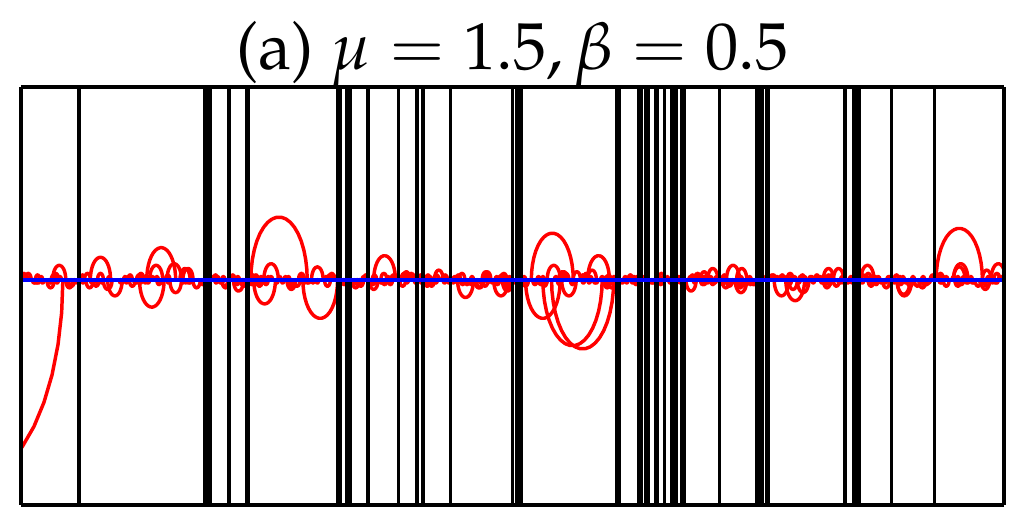}
\includegraphics[width=.9\columnwidth]{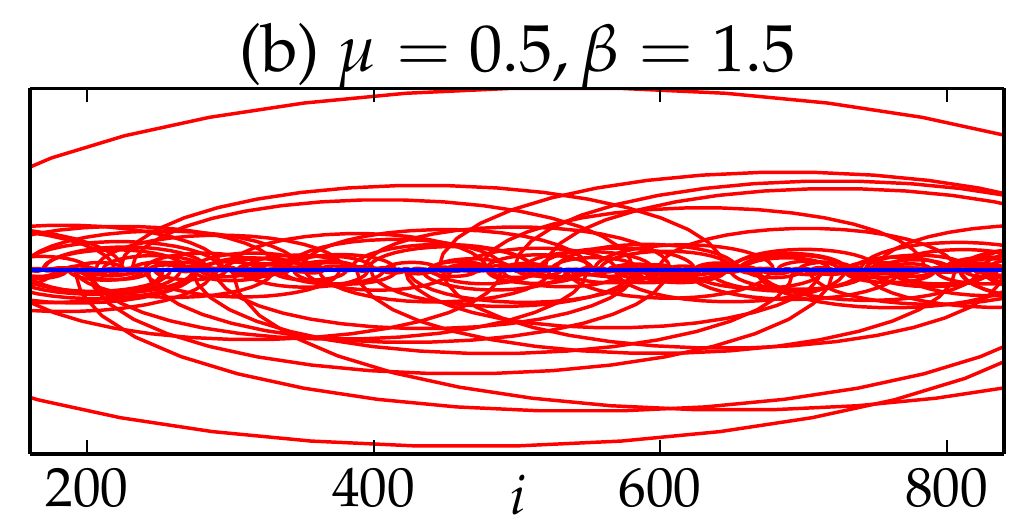}
\caption{Representative samples of the long--range percolation random graph $G$, see Eq. \eqref{eq:pmn}. The lattice is represented by the horizontal line, and the links by the semi--circles. The lattices have sites $i = 1, \dots, 1024$, with open boundary condition. (a) When $\mu > 1$, the graph is divided by the vertical lines (cuts) into intervals. (b) When $\mu < 1$, no cut is present.}\label{fig:lrp}
\end{figure}

The long--range percolation point of view also explains why there is no de--localisation in the $\mu > 1$ regime. For this, let us pick any site $i$, and divide the lattice into two halves: $\{n:n \leq i \}$ and $\{m: m \geq i+ 1 \}$. The probability that they are \textit{disconnected} in $G$ is 
\begin{equation} p_i = \prod_{m \leq i}\prod_{n \geq i} (1 - p_{mn}) \,. \label{eq:pi} \end{equation}
By Eq. \eqref{eq:pmn}, this infinite product is convergent if $\mu > 1$, and $p_i > 0$ if we exclude from the product the short--distance links present with $p_{mn} = 1$. As a consequence, $G$ is cut into a chain intervals of finite length such that adjacent intervals can be connected only by short--distance links, see Figure \ref{fig:lrp} (a) for illustration. Regarding each interval as one site, we obtain an effective 1d Anderson model, whose eigenstates are all localised. Note however that the size $s$ of the intervals is random and its distribution has an algebraic tail \footnote{Indeed, it is not hard to see that $\mathbb{P}(s > \abs{m-n}) \geq p_{mn} = \beta^{-1} \abs{m-n}^{-\mu-1}$.}; therefore, the localised states do not decay exponentially as in 1d Anderson models, but more slowly, as predicted by Eq. \eqref{eq:stretch}.

In contrast, as shown in Figure \ref{fig:lrp} (b),  when $\mu < 1$, the product Eq. \eqref{eq:pi} diverges to $0$, so no cut is present; correspondingly, there is a delocalisation transition, as shown in section \ref{sec:LT}. In this respect, the broadly distributed class is a ``quantum'' version of the long--range percolation model, in which the percolation transition is present (absent) for $\mu < 1$ ($\mu > 1$, respectively) \cite{schulman83lrp,newman1986lrp}.

As mentioned in section \ref{sec:num}, we can also predict the existence of mobility edges in the BBRM for arbitrarily large $\beta$, using a slightly different long--range percolation mapping, see appendix \ref{sec:LRP-ME}.

 \subsection{Effective dimension} \label{sec:dimension}
 The long--range percolation graph structure discussed above allows us to assign effective dimensions to the broadly distributed banded random matrices, by a comparison to Anderson models in $d$ dimension and on the Bethe lattice.
   
   For this, we consider the number of sites whose graph distance to a fixed site $0$ is less than $y$, denoted as $N(y)$. For the $d$--dimensional Anderson model, the graph in question is the $d$--dimensional lattice, and we have:
   \begin{equation}
   N(y) = C  y^d \,, \label{eq:euclid}
   \end{equation}
   where $C$ is a constant. For the Anderson model on a Bethe lattice with branching number $K$, and we have 
   \begin{equation}
   N(y) =  \exp (y \ln K) \,. \label{eq:cayley}
   \end{equation}
   We now consider $N(y)$ for the broadly distributed class. For this, note that $N(y)$ is the inverse function of the chemical distance $d_n$: $N(y) = n \Leftrightarrow d_n = y$, where we recall that $d_n \propto T_n$ [eq. \eqref{eq:dnTn}]. Then, Table \ref{table:summary} leads to the following:
   \begin{itemize}
   	\item[-] $\mu \geq 2 \Rightarrow N(y) \propto y$, giving an effective dimension $d=1$. 
   	\item[-] $\mu \in (1,2) \Rightarrow N(y) \propto y^{1/(\mu-1)}$,  giving an effective dimension $d=1/(\mu-1)$. As $\mu \searrow 1$, $d \nearrow \infty$. 
   	\item[-] $\mu \in (0,1) \Rightarrow N(y) \propto \exp(C y^{1/\kappa})$, with $\kappa > 1$: $N(y)$ grows faster than any finite dimension (Eq. \eqref{eq:euclid}), but slower than Cayley tree (Eq. \eqref{eq:cayley}). So the localisation transition and mobility edge in this regime are expected to be distinct from those in other mean-field models, such as  Anderson model on the Bethe lattice \cite{abou1973selfconsistent,abouchacra74self} and the model of Lévy matrices. A remarkable property of the latter is that the mobility edge can be analytically calculated. Doing the same for the broadly distributed class is a difficult challenge. 
   \end{itemize}
   
 \section{Conclusion}
In this work, we studied the broadly distributed random banded matrices. They are quasi--sparse long--range hopping models in 1d [section \ref{sec:definition}], and can be regarded as Anderson models on long--range percolation graphs [section \ref{sec:LRP}]. We demonstrated the existence of a localisation in the $\mu \in (0, 1)$ regime [section \ref{sec:LT}], and characterised the decay rates of the localised states [section \ref{sec:decay}], whose effective dimension interpolates between 1d and Bethe--lattice Anderson model [section \ref{sec:dimension}]. 

An intriguing question of timely interest is the connections of our model to many--body localisation. On one side, models with long--range many--body interactions display delocalisation thresholds exactly at $\mu = 0$ and $\mu = 1$ \cite{burin15mbl,burin2006energy,yao14mbldipolar}; it would be instructive to see whether deeper connections exist behind this coincidence. Another important aspect concerns the debate around the ``bad--metal'' phase \cite{Basko2006mbl}, in which the many--body eigenstates are multifractal in the Fock space. This space can be also seen as a graph, in which two sites are connected if the interaction matrix element between them is non--zero. Then, the number of sites at distance $\leq y$ from a given site increases faster than any power law $N(y) \gg y^n, \forall n$. For this reason, Anderson models on the Bethe lattice, regular random graphs, and the closely related Lévy matrices \cite{monthus16levy} were proposed as proxies of the Fock space. However, in all these models, $N(y)$ grows exponentially. Now, the growth of $N(y)$ is in fact slower than exponential for the Fock space. Indeed, a site in the Fock space has a fixed number of nearest neighbours, but many loops appear when considering next--nearest neighbours \textit{etc}, so that the exponential growth is an upper bound for $N(y)$.. In this respect, our model may be a better proxy in which the fate of the critical phase can be investigated.


\begin{acknowledgements} We thank E. Altman, A. Amir, G. Biroli, A. De Luca, F. Franchini, O. Giraud, A. Ludwig, C. Monthus, M. Ortu\~no, I. Rodriguez-Arias, C. Texier, T. Thiery and N. Yao for useful discussions. This work is supported by the Capital Fund Management Research Foundation, by “Investissements d’Avenir” LabEx PALM (ANR-10-LABX-0039-PALM), and ANR-16-CE30-0023-01 (THERMOLOC). We thank the hospitality of KITP (under Grant No. NSF PHY11-25915), where this work was initiated.
\end{acknowledgements}

 \appendix
 \section{Broad distribution and quasi--sparseness}\label{sec:math1}
  In this appendix, we show that the quasi--sparse property Eq. \eqref{eq:Qmndef0} can be entailed from the definition of the broadly distributed class, Eq. \eqref{eq:broaddefall}.
   For this we denote $Q = \abs{Q_{mn}} \geq 0$. Then the Laplace transform of $Q$ can be written as the cumulative distribution of $\ln Q$ convoluted with an random variable $\mathsf{Gum}$ drawn from the standard Gumbel distribution, independent of $Q$:
   \begin{align}
   \overline{\exp{(-t Q)}} &= \overline{\exp(-\exp(\ln Q + \ln t))} \nonumber \\
   & = \mathbb{P}(-\ln Q - \mathsf{Gum} > \ln t)
   \end{align} 
   Now Eq. \eqref{eq:broaddef} with $t = 1$ implies 
   \begin{equation}
   \mathbb{P}(-\ln Q - \mathsf{Gum} < \ln t) = f(t) / g + O(1/g^2) \,. \label{eq:cdf}
   \end{equation}
   Since $\mathsf{Gum}$ is of order unity, Eq. \eqref{eq:cdf} implies $Q \sim O(1)$ with probability $\propto 1/g$. 
   
   Now we look at the typical magnitude of $Q^{\text{typ}}$. It can be estimated by the value $\ln t$ for which the cumulative $\mathbb{P}(-\ln Q - G  < \ln t) = 1/2$, namely when $f(t) = g/2$ by Eq. \eqref{eq:cdf}. This gives $-\ln Q \sim y \sim \ln f^{\text{inv}}(g)$ and 
   \begin{equation} Q^{\text{typ}} \sim 1 / f^{\text{inv}}(g) < \exp(-cg) \,. \label{eq:Qtype}\end{equation} 
   where $f^{\text{inv}}$ is the inverse function of $f$ which grows at least exponentially according to Eq. \eqref{eq:ftlog}. So the only purpose of this technical condition is to guarantee that the typical elements are at most exponentially small.
 
 \section{Self--averaging property}\label{sec:selfave}
 In this appendix, we study the sum in Eq. \eqref{eq:defQtilde} in the simplest case where $c_i = 1$. This is a helpful warm--up for the more involved cases considered in section \ref{sec:block}, and illustrates the  relevance of the block--diagonalisation construction.
 
  Recall that for a broadly distributed matrix element $Q_{mn}$, the typical value is very small compared to its moments. On the other hand, since all the moments exist, for any fixed $g$, the central limit theorem applies. That is, if $Q_1, \dots,  Q_M$ are $M$ independent copies of $\abs{Q_{mn}}$, the sum
      \begin{equation}
      S \stackrel{\text{def}} = \sum_{i=1}^M Q_i  \label{eq:defS}
      \end{equation}
   tends to a Gaussian as $M\rightarrow\infty$ (after proper rescaling), whose moments and typical value are the same. When does the crossover happens? 
 
    The answer is $M \sim g$. In light of Eq. \eqref{eq:Qmndef0}, this is intuitive since $M \sim g$ is precisely when the rare event $Q_{mn} \sim O(1)$ begins to occur amongst $Q_1, \dots, Q_M$. Indeed, the distribution of $S$ has a well-defined limit $S_T$ when $M,g\rightarrow\infty$ with $M/g = T$ kept constant. To show this, recall the expansion $\overline{\exp(- t Q_{i})} = 1 - f(t) / g + O(1/g^2)$, Eq. \eqref{eq:broaddef}, which implies that:
    \begin{equation}
    \overline{\exp(-t S)} =  \left[\overline{\exp(- t Q_{i})}\right]^{M} \longrightarrow \exp(- T f(t))\,,\label{eq:limitdis}
    \end{equation}
    as $M = Tg \to \infty$. So, the distribution of $S$ has a limit $S_{T}$ depending on $T$, given in terms of Laplace transform $\overline{ \exp(-t S_{T}) }=  \exp(- T f(t))  \,. $  
    
    From Eq. \eqref{eq:limitdis} we conclude that when $M \sim g$, the distribution of the sum becomes $g$-independent in the $g\rightarrow\infty$ limit. Therefore, when $M \gg g$, we enter the central limit theorem regime, in which $S / \sqrt{M / g}$ tends to a Gaussian of variance of order $g$. On the other hand, when $M \ll g$, $T \ll 1$, Eq. \eqref{eq:limitdis} implies $\overline{ \exp(-t S_{T}) } \sim 1 - T f(t) + O(T^2)$, so the sum $S_T$ becomes itself broadly distributed, with $1 / T$ playing the rôle of $g$. 
    
  As an example, we note that for the BBRM, the cumulants of $S_T$ have a simple form  $ \overline{S_{T}^k}^c = T / k.$ Also, it is interesting to point out that the sum $S$ is identical to the partition function of the exponential random energy model \cite{bouchaud1997universality,bouchaud1998out}, yet with a different scaling of energy than the one that gives rise to a glassy transition.

\section{Resolvant and decay rate} \label{sec:resolvant}
In this appendix, we show how Eq. \eqref{eq:logresolvant} implies Eq. \eqref{eq:roTsccarsup}.

 For this, we write the operator $\hat{Q}$ in the basis of its eigenstates $\vert \lambda' \rangle$ of ${Q}$ (with energy $\lambda'$): 
 \begin{equation} \langle n \vert (1 - \hat{Q} / \lambda)^{-1} \vert 0 \rangle= \sum_{\lambda'} \frac{\langle n \vert \lambda' \rangle\langle \lambda' \vert 0 \rangle }{1 - \lambda'/\lambda} \,.  \label{eq:resolvantexpand} \end{equation} 
     Because of the denominator, the sum is dominated by eigenstates with energy close to $\lambda$, and the decay of those eigenstates determines the behaviour $\langle n \vert \lambda' \rangle\langle \lambda' \vert 0 \rangle$ as a function of $n$. Indeed, when $\vert\lambda'\rangle$ is localised around some site $m$ with decay $\ln \abs{\langle n \vert \lambda' \rangle} \propto -T_{\abs{m - n}}$, we have $ \ln \abs{\langle n \vert \lambda' \rangle\langle \lambda' \vert 0 \rangle }= - T_{\abs{m}} - T_{\abs{n-m}} \leq -T_{\abs{n}}$ (the last convexity inequality can be checked for all cases in Table \ref{table:summary}), so the eigenstates localised at $0$ or $n$ will dominate \eqref{eq:resolvantexpand} and give the same contribution:
     \begin{equation}
     \ln \abs{\langle n \vert (1 - \hat{Q} / \lambda)^{-1} \vert 0 \rangle} \approx \ln \abs{\langle n \vert \phi_0\rangle }  \,,
     \end{equation}
     where $\vert \phi_0\rangle$ is an eigenstate localised around $0$ having energy $\approx \lambda$. Combined with \eqref{eq:logresolvant}, we have

 \section{Long--range percolation argument for mobility edges}\label{sec:LRP-ME}
  In this appendix, we argue that the BBRM model in the $\mu \in (0, 1)$ regime has mobility edges when for arbitrarily large disorder, \textit{i.e.}, there is no upper--critical disorder $\beta_c$, see Figure \ref{fig:sparsephase}.
  
 The key of the argument is to define a resonance graph. For this, observe that the grand canonical partition function Eq. \eqref{eq:ZofBBRM2} is an infinite series that can diverge. Indeed, for any pair $m,m'$, the sub-series made of back--and--forth paths $(0, m,m',m,m',\dots,m,m',n)$, $\ell = 1,2,\dots$, which is
 $Q_{0m} Q_{m,m'}Q_{m'n}\lambda^{-3} \sum_{k=0}^{\infty} (Q_{mm'} / \lambda)^{2k}$, diverges when  
  \begin{equation} Q_{mn} > \lambda =  e^{-\beta \tau} \Leftrightarrow \tau_{mn} < \tau \,,\label{eq:resonancecondition} \end{equation}
 see Eq. \eqref{eq:ZofBBRM}. Such a divergence is associated with condensation phenomena in statistical physics. At the onset of divergence $Q_{mn} = \abs{\lambda} -\varepsilon$, the ``polymer'' tends to occupy infinite number of times on the link $m \leftrightarrow m'$.  Beyond that point, the statistical mechanics model is non--physical. However, from the quantum mechanics point of view, such divergence should be re-summed and interpreted as a \textit{resonance} between the sites $m$ and $m'$. From Eq. \eqref{eq:Qdef2}, one sees that the probability of such resonance is 
 \begin{equation} p_{nm} = 1 - e^{-\tau / \abs{n-m}^{\mu + 1}} \sim \frac{\tau}{\abs{n-m}^{\mu + 1}} \,, \end{equation}
 as $\abs{m-n}\gg 1$. We remark that, when $\beta \gg 1$, by Eq. \eqref{eq:rhobigb}, the resonance condition $Q_{mn} > \abs{\lambda} $ is equivalent to requiring that $Q_{mn}$ be larger than the energy level spacing.
 
 The random graph made of resonating edges defines thus another long--range percolation model.  When $\mu \in (0,1)$, there is a percolation threshold $\tau_c$ such that when $\tau > \tau_c$, the resonance graph is connected. Now, the percolation of the resonance graph is generally associated with the de-localisation of the eigenstate. Therefore, we expect that there are extended states with exponentially small eigenvalues $\abs{\lambda} < \lambda_c \approx e^{-\beta \tau_c}$ as $\beta\rightarrow\infty$. Since we know from section \ref{sec:block} that the eigenstates with $\lambda \sim O(1)$ (as $\beta \to \infty$) are localised, we expect mobility edges at:
 \begin{equation} \lambda_c \approx e^{-\beta \tau_c} \,, \beta \gg 1 \,. \label{eq:mobilityedge}\end{equation} 
 This argument is only qualitatively valid, because percolation does not necessarily imply de--localisation. For example, in 2D short--range lattices, percolation is possible, but the Anderson model does not have an extended phase (This is true for the original Anderson model; Anderson transitions in 2D are possible in a wider sense, see \cite{evers2008anderson} for a review). However, in our case, the estimate Eq. \eqref{eq:mobilityedge} captures qualitatively the phase diagram [Figure \ref{fig:sparsephase} (b)] that we observe numerically: the extended phase is indeed in the middle of the spectrum, and its size shrinks rapidly as $\beta$ increases.

  \bibliographystyle{apsrev4-1.bst}  
  \bibliography{rems}
 
 \end{document}